\documentclass[twocolumn,floatfix,eqsecnum,rmp]{revtex4}
\usepackage{epsfig}
\usepackage{amsmath}
\usepackage{bm}
\usepackage{bbold}
\newcommand{\sect}{\section}

\begin{document}

\title{Applications of random matrix theory\\
to condensed matter and optical physics}
\author{C.W.J. Beenakker}
\affiliation{Instituut-Lorentz, Universiteit Leiden, P.O. Box 9506, 2300 RA Leiden, The Netherlands}
\begin{abstract}
{\tt Two chapters for \textit{The Oxford Handbook of Random Matrix Theory}, edited by G. Akemann, J. Baik, and P. Di Francesco, to be published by Oxford University Press.}
\end{abstract}
\maketitle
\tableofcontents

%%%%%%%%%%%%%%%%%%%%%%%%%%%%%%%%%%%%%%%%%%%%%%%%%%%%%%%%%%%%%%%%%%%%%%%%%%%

\part{Condensed Matter Physics}
\label{CMP}

\sect{Introduction}\label{intro_condmat}

Applications of random matrix theory (RMT) to condensed matter physics search for universal features in the electronic properties of metals, originating from the universality of eigenvalue repulsion. Eigenvalue repulsion is universal, because the Jacobian
\begin{equation}
J=\prod_{i<j}|E_{j}-E_{i}|^{\beta}\label{Jdef}
\end{equation} 
of the transformation from matrix space to eigenvalue space depends on the symmetry of
the random matrix ensemble (expressed by the index $\beta\in\{1,2,4\}$) --- but is independent of microscopic properties such as the mean eigenvalue separation \cite{Meh67}. This universality is at the origin of the remarkable success of RMT in nuclear physics \cite{Bro81,Wei09}.

In condensed matter physics, the applications of RMT fall into two broad categories. In the first category, one studies thermodynamic properties of closed systems, such as metal grains or semiconductor quantum dots. The random matrix is the Hamiltonian $H$. In the second category, one studies transport properties of open systems, such as metal wires or quantum dots with point contacts. Now the random matrix is the scattering matrix $S$ (or a submatrix, the transmission matrix $t$). Applications in both categories have flourished with the development of nanotechnology. Confinement of electrons on the nanoscale in wire geometries (quantum wires) and box geometries (quantum dots) preserves their phase coherence, which is needed for RMT to be applicable.

The range of electronic properties addressed by RMT is quite broad. The selection of topics presented in this Chapter is guided by the desire to show those applications of RMT that have actually made an impact on experiments. For a more complete coverage of topics and a more comprehensive list of references we suggest a few review articles \cite{Bee97,Guh98,Alh00}.

\sect{Quantum wires}\label{wires}
\subsection{Conductance fluctuations}\label{UCF_sec}

In the 1960's, Wigner, Dyson, Mehta, and others discovered that the fluctuations in the energy level density are governed by level repulsion and therefore take a universal form \cite{Por65}.
The universality of the level fluctuations is expressed by the
Dyson-Mehta formula \cite{Dys63} for the variance of a linear statistic\footnote{
The quantity $A$ is
called a linear statistic because products of different $E_{n}$'s do
not appear, but the function $a(E)$ may well depend non-linearly on
$E$.}
$A=\sum_{n}a(E_{n})$ on the energy levels $E_{n}$.  The Dyson-Mehta formula reads 
\begin{equation}
{\rm Var}\,A=\frac{1}{\beta}\,\frac{1}
{\pi^{2}}\int_{0}^{\infty}dk\,|a(k)|^{2}k,\label{DysonMehta}
\end{equation}
where $a(k)=\int_{-\infty}^{\infty}\!dE\,{\rm e}^{{\rm i}kE}a(E)$ is
the Fourier transform of $a(E)$.  Eq.\ \eqref{DysonMehta} shows that:
1.~The variance is independent of microscopic parameters; 2.~The
variance has a universal $1/\beta$-dependence on the symmetry index.

\begin{figure}[!tb]
\unitlength1cm
\centerline{
\epsfig{figure=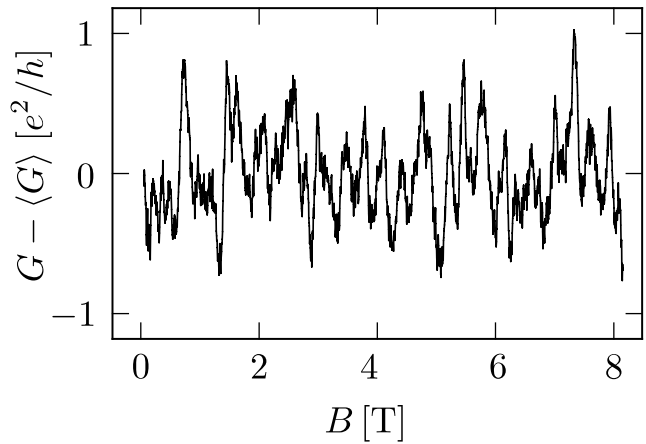,width=20pc}
}
\caption{
Fluctuations as a function of perpendicular magnetic field of the conductance of a 310~nm long and 25~nm wide Au wire at 10~mK. The trace appears random, but is completely reproducible from one measurement to the next. The root-mean-square of the fluctuations is $0.3\,e^{2}/h$, which is not far from the theoretical result $\sqrt{1/15}\,e^{2}/h$ [Eq.\ \eqref{UCF} with $\beta=2$ due to the magnetic field and a reduced conductance quantum $G_{0}=e^{2}/h$ due to the strong spin-orbit scattering in Au]. Adapted from Ref.\ \cite{Was86}.
\label{fig_UCFAu}}
\end{figure}

In a pair of seminal 1986-papers \cite{Imr86,Alt86}, Imry and Altshuler and Shklovkski\u{\i} proposed to apply RMT to the
phenomenon of universal conductance fluctuations (UCF) in metals, which was discovered
using diagrammatic perturbation theory by Altshuler \cite{Alt85} and
Lee and Stone \cite{Lee85}. UCF is the occurrence of sample-to-sample
fluctuations in the conductance which are of order $e^{2}/h$ at zero
temperature, \textit{independent} of the size of the sample or the
degree of disorder --- as long as the conductor remains in the diffusive
metallic regime (size $L$ large compared to the mean free path $l$, but small compared to the localization length $\xi$). An example is shown in Fig.\ \ref{fig_UCFAu}. 

The similarity between the statistics of energy
levels measured in nuclear reactions on the one hand, and the
statistics of conductance fluctuations measured in transport
experiments on the other hand, was used by Stone \textit{et al.} \cite{Mut87,Sto91} to construct a random matrix theory of quantum
transport in metal wires. The random matrix is now not the Hamiltonian $H$, but the transmission matrix $t$, which determines the conductance through the Landauer formula
\begin{equation}
G=G_{0}{\rm Tr}\,tt^{\dagger}=G_{0}\sum_{n}T_{n}.\label{Landauer}
\end{equation}
The conductance quantum is $G_{0}=2e^{2}/h$, with a factor of two to account for spin degeneracy. Instead of repulsion of energy levels, one now has repulsion of the transmission eigenvalues $T_{n}$, which are the eigenvalues of the transmission matrix product $tt^{\dagger}$. In a wire of cross-sectional area ${\cal A}$ and Fermi wave length $\lambda_{F}$, there are of order $N\simeq A/\lambda_{F}^{2}$ propagating modes, so $t$ has dimension $N\times N$ and there are $N$ transmission eigenvalues. The phenomenon of UCF applies to the regime $N\gg 1$, typical for metal wires.

Random matrix theory is based on the fundamental
assumption that all correlations between the eigenvalues are due to the
Jacobian  $J=\prod_{i<j}|T_{i}-T_{j}|^{\beta}$ from matrix elements to eigenvalues. If all correlations are due to the Jacobian, then the
probability distribution $P(T_{1},T_{2},\ldots T_{N})$
of the $T_{n}$'s should have the form $P\propto
J\prod_{i}p(T_{i})$, or equivalently,
\begin{align}
P(\{T_{n}\})\propto{}&\exp\Bigl[-\beta\Bigl(\sum_{i<j}
u(T_{i},T_{j})+\sum_{i}V(T_{i})\Bigr)\Bigr],
\label{Pglobala}\\
u(T_{i},T_{j})={}&-\ln|T_{j}-T_{i}|,
\label{Pglobalb}
\end{align}
with $V=-\beta^{-1}\ln p$.
Eq.\ \eqref{Pglobala} has the form of a Gibbs distribution at
temperature $\beta^{-1}$ for a fictitious system of classical particles
on a line in an external potential $V$, with a logarithmically
repulsive interaction $u$. All microscopic parameters are contained in
the single function $V(T)$. The logarithmic repulsion is
independent of microscopic parameters, because of its geometric
origin.

Unlike the RMT of energy levels, the correlation function of the $T_{n}$'s is not
translationally invariant, due to the
constraint $0\leq T_{n}\leq 1$ imposed by unitarity of the scattering matrix. Because of this
constraint, the Dyson-Mehta formula \eqref{DysonMehta} needs to be modified, as shown in Ref.\ \cite{Bee93a}. In the large-$N$ limit, the variance of a linear
statistic $A=\sum_{n}f(T_{n})$ on the transmission eigenvalues is given by
\begin{equation}
{\rm Var}\,A=\frac{1}{\beta}\,\frac{1}{\pi^{2}}\int_{0}^{\infty}
dk\,|F(k)|^{2}k\tanh(\pi k).\label{CB}
\end{equation}
The function $F(k)$ is defined in terms of the function $f(T)$ by the
transform
\begin{equation}
F(k)=\int_{-\infty}^{\infty}\!dx\,{\rm e}^{{\rm i}kx}f
\left(\frac{1}{1+{\rm e}^{x}}\right).\label{Fkdef}
\end{equation}
The formula \eqref{CB} demonstrates that the universality which was the
hallmark of UCF is generic for a whole class of transport properties,
viz.\ those which are linear statistics on the transmission
eigenvalues. Examples, reviewed in Ref.\ \cite{Bee97}, are the critical-current fluctuations in Josephson
junctions, conductance fluctuations at normal-superconductor interfaces, and fluctuations in the
shot-noise power of metals.

\subsection{Nonlogarithmic eigenvalue repulsion}\label{nonlog}

The probability distribution \eqref{Pglobala} was justified by a
maximum-entropy principle for an ensemble of quasi-1D conductors \cite{Mut87,Sto91}.
Quasi-1D refers to a wire geometry (length $L$ much greater than width $W$). In such a geometry one can assume
that the distribution of scattering matrices in an ensemble with different realizations of the disorder is only a function of the
transmission eigenvalues (isotropy assumption). The distribution
\eqref{Pglobala} then maximizes the information entropy subject to the
constraint of a given density of eigenvalues. The function $V(T)$
is determined by this constraint and is not specified by RMT.

It was initially believed that Eq.\ \eqref{Pglobala} would provide an
exact description in the quasi-1D limit $L\gg W$, if only $V(T)$ were
suitably chosen \cite{Sto91}. However, the generalized Dyson-Mehta
formula (\ref{CB}) demonstrates that RMT is not exact in a quantum wire \cite{Bee93a}. If one computes from Eq.\ \eqref{CB} the variance of
the conductance (\ref{Landauer}) [by substituting $f(T)=G_{0}T$], one finds
\begin{equation}
{\rm Var\,}G/G_{0}=\frac{1}{8}\beta^{-1},\label{UCFglobal}
\end{equation}
independent of the form of $V(T)$. The diagrammatic perturbation
theory \cite{Alt85,Lee85} of UCF gives instead
\begin{equation}
{\rm Var\,}G/G_{0}=\frac{2}{15}\beta^{-1}\label{UCF}
\end{equation}
for a quasi-1D conductor. The difference between the coefficients
$\frac{1}{8}$ and $\frac{2}{15}$ is tiny, but it has the fundamental
implication that the interaction between the $T$'s is not
precisely logarithmic, or in other words, that there exist correlations
between the transmission eigenvalues over and above those induced by
the Jacobian \cite{Bee93a}.

The $\frac{1}{8}$~---~$\frac{2}{15}$ discrepancy raised the question
what the true eigenvalue interaction would be in quasi-1D conductors.
Is there perhaps a cutoff for large separation of the $T$'s? Or
is the true interaction a many-body interaction, which cannot be
reduced to the sum of pairwise interactions? This transport problem has
a counterpart in a closed system. The RMT of the statistics of the
eigenvalues of a random Hamiltonian yields a probability distribution
of the form \eqref{Pglobala}, with a logarithmic repulsion between the
energy levels \cite{Meh67}. It was shown by Efetov \cite{Efe83} and by
Altshuler and Shklovski\u{\i} \cite{Alt86} that the logarithmic
repulsion in a disordered metal grain holds for energy separations small compared to the inverse ergodic time $\hbar/\tau_{\rm erg}$.\footnote{
The ergodic time is the time needed for a particle to explore the available phase space in a closed system. In a disordered metal grain of size $L$ and diffusion constant $D$, one has $\tau_{\rm erg}\simeq L^{2}/D$. If the motion is ballistic (with velocity $v_{F}$) rather than diffusive, one has instead $\tau_{\rm erg}\simeq L/v_{F}$.}
For larger separations
the interaction potential decays algebraically \cite{Jal93}. 

The way in which the RMT of quantum transport breaks down is
quite different \cite{Bee93b}. The probability distribution of the transmission eigenvalues does indeed take the form \eqref{Pglobala} of a Gibbs distribution with a parameter-independent two-body interaction
$u(T_{i},T_{j})$, as predicted by RMT. However, the
interaction differs from the logarithmic repulsion \eqref{Pglobalb} of
RMT. Instead, it is given by
%reformatted
\begin{align}
u(T_{i},T_{j})={}&-{\textstyle\frac{1}{2}}
\ln|T_{j}-T_{i}|
-{\textstyle\frac{1}{2}}\ln|x_{j}-x_{i}|,\nonumber\\
&{\rm with}\;\;T_{n}\equiv 1/\cosh^{2}x_{n}.\label{final3u}
\end{align}
The eigenvalue interaction \eqref{final3u} is different for weakly and
for strongly transmitting scattering channels:
$u\rightarrow-\ln|T_{j}-T_{i}|$ for
$T_{i},T_{j}\rightarrow 1$, but
$u\rightarrow-\frac{1}{2}\ln|T_{j}-T_{i}|$ for
$T_{i},T_{j}\ll 1$. For weakly transmitting channels it is
\textit{twice as small} as predicted by considerations based solely on
the Jacobian, which turn out to apply only to the strongly transmitting
channels.

The nonlogarithmic interaction modifies the Dyson-Mehta formula for the variance of a linear statistic. Instead of Eq.\ \eqref{CB} one now has \cite{Bee93b,Cha93}
\begin{align}
&{\rm Var\,}A=\frac{1}{\beta}\,\frac{1}{\pi^{2}}\int_{0}^{\infty}dk\,
\frac{k|F(k)|^{2}}{1+{\rm cotanh}
({\textstyle\frac{1}{2}}\pi k)},\label{VarAresult}\\
&F(k)=\int_{-\infty}^{\infty}dx\,{\rm e}^{{\rm i}kx}
f\left(\frac{1}{\cosh^{2}x}\right).\label{akdef}
\end{align}
Substitution of $f(T)=T$ now yields $\frac{2}{15}$ instead of $\frac{1}{8}$ for the
coefficient of the UCF, thus resolving the discrepancy between
Eqs.\ \eqref{UCFglobal} and \eqref{UCF}.

The result \eqref{final3u} follows from the solution of a differential equation which determines how the probability distribution of the $T_{n}$'s changes when the length $L$ of the wire is incremented. This differential equation has the form of a multivariate drift-diffusion equation (with $L$ playing the role of time) for $N$ classical particles at coordinates $x_{n}={\rm arcosh}\,T_{n}^{-1/2}$. The drift-diffusion equation,
%reformatted
\begin{widetext}
\begin{align}
&l\frac{\partial }{\partial L}P(\{x_{n}\},L)=
\frac{1}{2}(\beta N+2-\beta)^{-1}\sum_{n=1}^{N}\frac{\partial}{\partial x_{n}}
\Bigl(\frac{\partial P}{\partial x_{n}}+
\beta P\frac{\partial\Omega}{\partial x_{n}}\Bigr),
\label{FPxa}\\
&\Omega=-\sum_{i=1}^{N}\sum_{j=i+1}^{N} \ln|\sinh^{2}x_{j}-\sinh^{2}x_{i}|-\frac{1}{\beta}\sum_{i=1}^{N}\ln|\sinh 2x_{i}|,\label{FPxb}
\end{align}
\end{widetext}
is known as the DMPK equation, after the scientists who first studied its properties in the 1980's \cite{Dor82,Mel88a}. (The equation itself appeared a decade earlier \cite{Bur72}.) The DMPK equation can be solved exactly \cite{Bee93b,Cas95}, providing the nonlogarithmic repulsion \eqref{final3u}.

\subsection{Sub-Poissonian shot noise}\label{noise}

The average transmission probability $\bar{T}=l/L$ for diffusion through a wire is the ratio of mean free path $l$ and wire length $L$. This average is not representative for a single transmission eigenvalue, because eigenvalue repulsion prevents the $T_{n}$'s from having a narrow distribution around $\bar{T}$. The eigenvalue density $\rho(T)=\langle\sum_{n}\delta(T-T_{n})\rangle$ can be calculated from the DMPK equation \eqref{FPxa}, with the result \cite{Dor84,Mel89}
\begin{equation}
\rho(T)=\frac{Nl}{2L}\frac{1}{T\sqrt{1-T}},\;\;{\rm for}\;\;T_{\rm min}\leq T<1,\label{rhoresult}
\end{equation}
in the diffusive metallic regime\footnote{
The localization length $\xi$ also follows from the DMPK equation. It is given by $\xi=(\beta N+2-\beta)l$, so it is larger than $l$ by a factor of order $N$.}
$l\ll L\ll\xi$. The lower limit $T_{\rm min}$ is determined by the normalization, $\int_{0}^{1} dT\,\rho(T)=N$, giving $T_{\rm min}\approx 4e^{-L/2l}$ with exponential accuracy. 

The transmission eigenvalue density is \textit{bimodal}, with a peak at unit transmission (open channels) and a peak at exponentially small transmission (closed channels). This bimodal distribution cannot be observed in the conductance $G\propto\sum_{n}T_{n}$, which would be the same if all $T_{n}$'s would cluster near the average $\bar{T}$. The shot noise power ${\cal S}\propto\sum_{n}T_{n}(1-T_{n})$ (the second moment of the time dependent current fluctuations) provides more information. 

\begin{figure}[!tb]
\unitlength1cm
\centerline{
\epsfig{figure=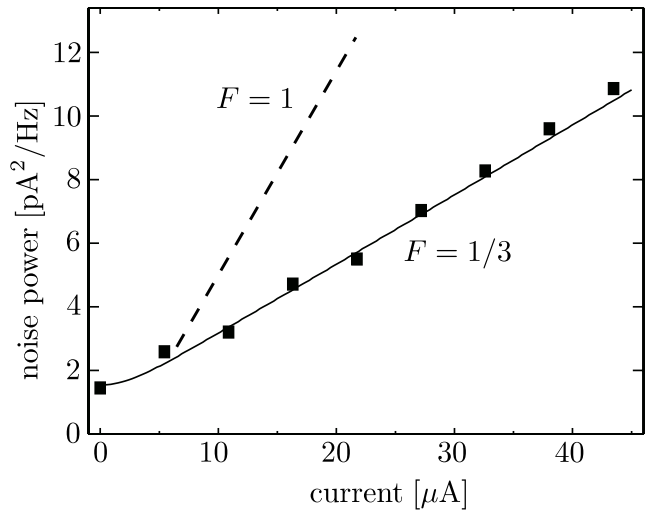,width=20pc}
}
\caption{
Sub-Poissonian shot noise in a disordered gold wire (dimensions $940\,{\rm nm} \times 100\,{\rm nm}$). At low currents the noise saturates at the level set by the temperature of $0.3\,{\rm K}$. At higher currents the noise increases linearly with current, with a slope that is three times smaller than the value expected for Poisson statistics. Adapted from Ref.\ \cite{Hen99}.
\label{fig_henny}}
\end{figure}

The ratio of shot noise power and conductance, defined in dimensionless form by the Fano factor 
\begin{equation}
F=\frac{\sum_{n}T_{n}(1-T_{n})}{\sum_{n}T_{n}},\label{Fdef}
\end{equation}
quantifies the deviation of the current fluctuations from a Poisson process (which would have $F=1$). Since $\bar{T}\ll 1$, if all $T_{n}$'s would be near $\bar{T}$ the current fluctuations would have Poisson statistics with $F=1$. The bimodal distribution \eqref{rhoresult} instead gives sub-Poissonian shot noise \cite{Bee92},
\begin{equation}
F\rightarrow 1-\frac{\int dT\,\rho(T)T^{2}}{\int dT\,\rho(T)T}=1-\frac{2}{3}=\frac{1}{3}.\label{Fonethird}
\end{equation}
(The replacement of the sum over $n$ by an integration over $T$ with weight $\rho(T)$ is justified in the large-$N$ limit.) This one-third suppression of shot noise below the Poisson value has been confirmed experimentally \cite{Ste96,Hen99}, see Fig.\ \ref{fig_henny}.

\sect{Quantum dots}\label{dots}

\subsection{Level and wave function statistics}\label{levelwave}

%reformatted
\begin{figure*}[!tb]
\unitlength1cm
\centerline{
\epsfig{figure=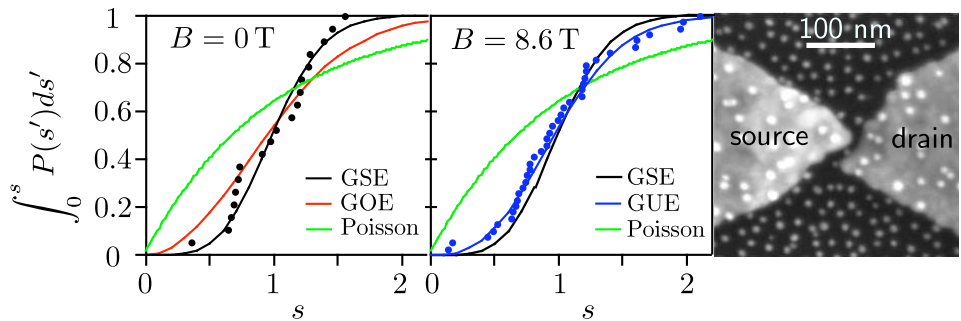,width=30pc}
}
\caption{
Data points: Integrated level spacing distribution of a single 10~nm diameter gold particle (barely visible in the micrograph as a white dot touching source and drain electrodes), measured by resonant tunneling in zero magnetic field and in a high magnetic field. The level spacings $s$ are normalized by the mean level spacing $\delta$ (equal to 0.23~meV in zero field and reduced to 0.12~meV in high fields due to splitting of the spin degenerate levels by the Zeeman effect). The measured distributions are compared with Wigner's RMT prediction: $P(s)\propto s^{\beta}\exp(-c_{\beta}s^{2})$ (with $c_{1}=\pi/4$, $c_{2}=4/\pi$, $c_{4}=64/9\pi$), for the Gaussian orthogonal ensemble (GOE, $\beta=1$), the Gaussian unitary ensemble (GUE $\beta=2$) and the Gaussian symplectic ensemble (GSE, $\beta=4$). The Poisson distribution $P(s)\propto e^{-s}$ of uncorrelated levels is also shown. A magnetic field causes a transition from the symplectic ensemble in zero field (preserved time reversal symmetry, broken spin rotation symmetry due to the strong spin-orbit coupling in gold), to the unitary ensemble in high fields (broken time reversal and spin rotation symmetries). Adapted from Ref.\ \cite{Kue08}.
\label{fig_spacing_grain}}
\end{figure*}

Early applications of random matrix theory to condensed matter physics were due to Gorkov and Eliashberg \cite{Gor65} and to Denton, M\"{u}hlschlegel, and Scalapino \cite{Den71}. They took the Gaussian orthogonal ensemble to model the energy level statistics of small metal grains and used it to calculate quantum size effects on their thermodynamic properties. (See Ref.\ \cite{Hal86} for a review.) Theoretical justification came with the supersymmetric field theory of Efetov \cite{Efe83}, who derived the level correlation functions in an ensemble of disordered metal grains and showed that they agree with the RMT prediction up to an energy scale of the order of the inverse ergodic time $\hbar/\tau_{\rm erg}$.

Experimental evidence for RMT remained rare throughout the 1980's --- basically because the energy resolution needed to probe spectral statistics on the scale of the level spacing was difficult to reach in metal grains. Two parallel advances in nanofabrication changed the situation in the 1990's. 

One the one hand, it became possible to make electrical contact to individual metal particles of diameters as small as 10~nm \cite{Del01}. Resonant tunneling through a single particle could probe the energy level spectrum with sufficient accuracy to test the RMT predictions \cite{Kue08} (see Fig.\ \ref{fig_spacing_grain}). 

On the other hand, semiconductor quantum dots became available. A quantum dot is a cavity of sub-micron dimensions, etched in a semiconducting two-dimensional electron gas. The electron wave length $\lambda_{F}\simeq 50\,{\rm nm}$ at the Fermi energy in a quantum dot is two order of magnitudes greater than in a metal, and the correspondingly larger level spacing makes these systems ideal for the study of spectral statistics. The quantum dot may be disordered (mean free path $l$ less than its linear dimension $L$) or it may be ballistic ($l$ greater than $L$). RMT applies on energy scales $\hbar/\tau_{\rm erg}\simeq(\hbar v_{F}/L)\min(1,l/L)$ irrespective of the ratio of $l$ and $L$, provided that the classical dynamics is chaotic.

Resonant tunneling through quantum dots has provided detailed information on both the level spacing distribution (through the spacing of the resonances) and on the wave function statistics (through the peak height of the resonances) \cite{Alh00}. For resonant tunneling through single-channel point contacts (tunnel probability $\Gamma$) the conductance peak height $G_{\rm max}$ is related to the wave function intensities $I_{1}$, $I_{2}$ at the two point contacts by \cite{Bee91}
\begin{equation}
G_{\rm max}=\frac{e^{2}}{h}\frac{\Gamma\delta}{4k_{B}T}\frac{I_{1}I_{2}}{I_{1}+I_{2}}.\label{Gmax}
\end{equation}
(The intensities are normalized to unit average and $\delta$ is the mean energy level spacing. The thermal energy $k_{B}T$ is assumed to be large compared to the width $\Gamma\delta$ of the resonances but small compared to $\delta$.) 

The Porter-Thomas distribution $P(I)\propto I^{\beta/2-1}e^{-\beta I/2}$ of (independently fluctuating) intensities $I_{1},I_{2}$ in the GOE ($\beta=1$) and GUE ($\beta=2)$ then gives the peak height distribution \cite{Jal92,Pri93},
\begin{equation}
P(g)=\left\{\begin{array}{ll}
(\pi g)^{-1/2}e^{-g},&\beta=1,\\
g[K_{0}(g)+K_{1}(g)]e^{-g},&\beta=2,
\end{array}\right.\label{Pgresult}
\end{equation}
with $g=(8k_{B}T/\Gamma\delta)(h/e^{2})G_{\rm max}$ and Bessel functions $K_{0}$, $K_{1}$. A comparison of this RMT prediction with available experimental data has shown a consistent agreement, with some deviations remaining that can be explained by finite-temperature effects and effects of exchange interaction \cite{Alh02}.

\subsection{Scattering matrix ensembles}\label{Gausstocirc}

In quantum dots, the most comprehensive test of RMT has been obtained by studying the statistics of the scattering matrix $S$ rather than of the Hamiltonian $H$. The Hamiltonian $H$ and scattering matrix $S$ of a quantum dot are related by \cite{Bla91,Got08}
\begin{align}
S(E)&={\mathbb 1}-2\pi iW^{\dagger}(E-H+i\pi WW^{\dagger})^{-1}W\nonumber\\
&=\frac{{\mathbb 1}+i\pi W^{\dagger}(H-E)^{-1}W}{{\mathbb 1}-i\pi W^{\dagger}(H-E)^{-1}W}.\label{SHeq}
\end{align}
The $M\times (N_{1}+N_{2})$ coupling matrix $W$ (assumed to be independent of the energy $E$) couples the $M$ energy levels in the quantum dot to $N_{1}+N_{2}$ scattering channels in a pair of point contacts that connect the quantum dot to electron reservoirs. The eigenvalue $w_{n}$ of the coupling-matrix product $W^{\dagger}W$ is related to the transmission probability $\Gamma_{n}\in[0,1]$ of mode $n$ through the point contact by
\begin{equation}
\Gamma_{n}=\frac{4\pi^{2}w_{n}M\delta}{(M\delta+\pi^{2}w_{n})^{2}}.\label{w2Gammaa}
\end{equation}
Eq.\ \eqref{SHeq} is called the Weidenm\"{u}ller formula in the theory of chaotic scattering, because of pioneering work by Hans Weidenm\"{u}ller and his group \cite{Mah69}.

A distribution function $P(H)$ for the Hamiltonian $H$ implies a distribution \textit{functional} $P[S(E)]$ for the scattering matrix $S(E)$. For electrical conduction at low voltages and low temperatures, the energy may be fixed at the Fermi energy $E_{F}$ and knowledge of the distribution function $P(S_{0})$ of $S_{0}=S(E_{F})$ is sufficient. For the Hamiltonian we take the Gaussian ensemble,
\begin{equation}
P(H)\propto\exp\left(-\beta(\pi/2\delta)^{2}M^{-1}\,{\rm Tr}\,H^2\right),\label{GaussEns}
\end{equation}
and we take the limit $M\rightarrow\infty$ (at fixed $\delta$, $E_{F}$, $\Gamma_{n}$), appropriate for a quantum dot of size $L\gg\lambda_{F}$. The number of  channels $N_{1},N_{2}$ in the two point contacts may be as small as $1$, since the opening of the point contacts is typically of the same order as $\lambda_{F}$. 

As derived by Brouwer \cite{Bro95}, Eqs.\ \eqref{SHeq} and \eqref{GaussEns} together imply, in the large-$M$ limit, for $S_{0}$ a distribution of the form
\begin{equation}
P(S_{0})\propto|{\rm Det}({\mathbb 1}-\bar{S}^{\dagger}S_{0})|^{-\beta N_{1}-\beta N_{2}-2+\beta}, \label{Poissonkernel2}
\end{equation}
known as the \textit{Poisson kernel} \cite{Hua63,Lew91,Dor92}. The average scattering matrix\footnote{
The average $\bar{S}$ is defined by integration over the unitary group with Haar measure $dS_{0}$, unconstrained for $\beta=2$ and subject to the constraints of time reversal symmetry for $\beta=1$ (when $S$ is symmetric) or symplectic symmetry for $\beta=4$ (when $S$ is self-dual). For more information on integration over the unitary group, see Refs.\ \cite{Bee97,Guh98}.
}
$\bar{S}=\int dS_{0}\,S_{0}P(S_{0})$ in the Poisson kernel is given by
\begin{equation}
\bar{S}=\frac{M\delta-\pi^{2}W^{\dagger}W}{M\delta+\pi^{2}W^{\dagger}W}.\label{SbarSupSym2}
\end{equation}

The case of ideal coupling (all $\Gamma_{n}$'s equal to unity) is of particular interest, since it applies to the experimentally relevant case of ballistic point contacts (no tunnel barrier separating the quantum dot from the electron reservoirs). In view of Eq.\ \eqref{w2Gammaa} one then has $\bar{S}=0$, hence
\begin{equation}
P(S_{0})={\rm constant}.\label{PS0}
\end{equation}
This is the distribution of Dyson's circular ensemble \cite{Dys62}, first applied to quantum scattering by Bl\"{u}mel and Smilansky \cite{Blu90}. 

The circular ensemble of scattering matrices implies for the $\min(N_{1},N_{2})$ nonzero transmission eigenvalues the distribution \cite{Bee97}
\begin{equation}
P(\{T_{n}\})\propto\prod_{n<m}|T_{n}-T_{m}|^{\beta}\prod_{k}T_{k}^{\frac{1}{2} \beta(|N_{2}-N_{1}|+1-2/\beta)}. \label{PTcircular}
\end{equation}
This distribution is of the form \eqref{Pglobala}, with the logarithmic repulsion \eqref{Pglobalb}. There are no nonlogarithmic corrections in a quantum dot, unlike in a quantum wire.

\subsection{Conductance distribution}\label{UCF_qd}

The complete probability distribution of the conductance $G=G_{0}\sum_{n=1}^{\min(N_{1},N_{2})}T_{n}$ follows directly from Eq.\ \eqref{PTcircular} in the case $N_{1}=N_{2}=1$ of single-channel ballistic point contacts \cite{Bar94,Jal94},
\begin{equation}
P(G)=\frac{\beta}{2G_{0}}(G/G_{0})^{-1+\beta/2},\;\;0<G<G_{0}. \label{PTN1Gamma0}
\end{equation}
This strongly non-Gaussian distribution rapidly approaches a Gaussian with increasing $N_{1}=N_{2}\equiv N$. Experiments typically find a conductance distribution which is closer to a Gaussian even in the single-channel case \cite{Hui98}, due to thermal averaging and loss of phase coherence at finite temperatures.

\begin{figure}[!tb]
\unitlength1cm
\centerline{
\epsfig{figure=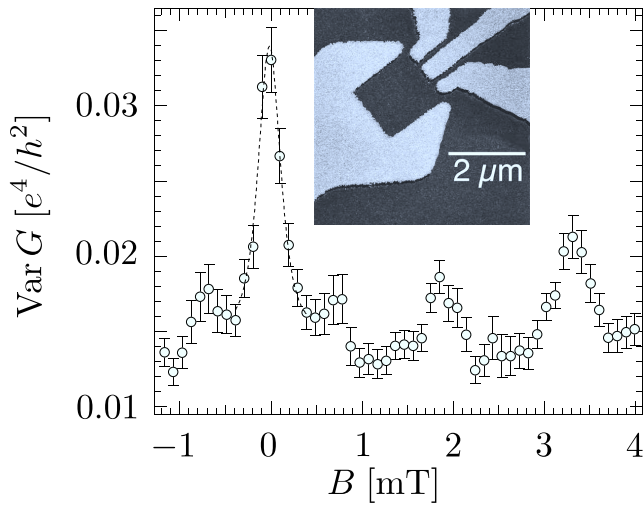,width=20pc}
}
\caption{
Variance of the conductance of a quantum dot at $30\,{\rm mK}$, as a function of magnetic field. The inset shows an electron micrograph of the device, fabricated in the two-dimensional electron gas of a GaAs/AlGaAs heterostructure. The black rectangle at the center of the inset is the quantum dot, the gray regions are the gate electrodes on top of the heterostructure. Electrons can enter and exit the quantum dot through point contacts at the top and right corner of the rectangle. The side of the rectangle between these two corners is distorted to generate conductance fluctuations and obtain the variance. Adapted from Ref.\ \cite{Cha95}.
\label{fig_qd}}
\end{figure} 

In the limit $N\rightarrow\infty$ the variance of the Gaussian is given by the RMT result \eqref{UCFglobal} for UCF --- without any corrections since the eigenvalue repulsion in a quantum dot is strictly logarithmic. The experiment value in Fig.\ \ref{fig_qd} is smaller than this zero-temperature result, but the factor-of-two reduction upon application of a magnetic field ($\beta=1\rightarrow\beta=2$) is quite well preserved.

\begin{figure}[!tb]
\unitlength1cm
\centerline{
\epsfig{figure=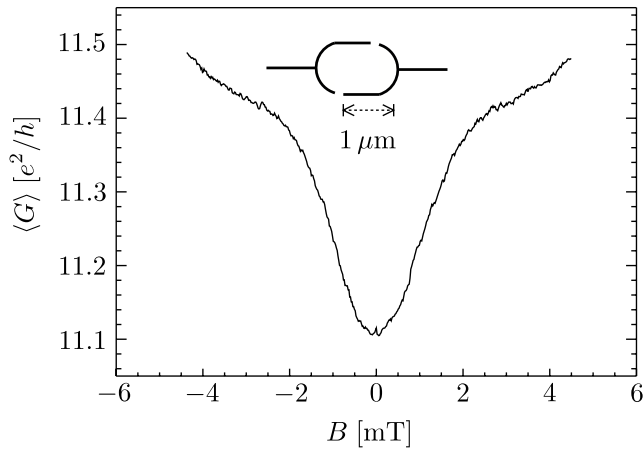,width=20pc}
}
\caption{
Magnetoconductance at $50\,{\rm mK}$, averaged over 48 quantum dots. The minimum around zero magnetic field is the weak localization effect. The inset shows the geometry of the quantum dots, which are fabricated in the two-dimensional electron gas of a GaAs/AlGaAs heterostructure. Adapted from Ref.\ \cite{Cha94}.
\label{fig_Chang}}
\end{figure}

Without phase coherence the conductance would have average $G_{0}N/2$, corresponding to two $N$-mode point contacts in series. Quantum interference corrects that average, $\langle G\rangle=G_{0}N/2+\delta G$. The correction $\delta G$ in the limit $N\rightarrow\infty$, following from the circular ensemble, equals
\begin{equation}
\delta G=\frac{1}{4}\left(1-\frac{2}{\beta}\right)G_{0}.\label{deltaG}
\end{equation}
The quantum correction vanishes in the presence of a time-reversal-symmetry breaking magnetic field ($\beta=2$), while in zero magnetic field the correction can be negative ($\beta=1$) or positive ($\beta=4$) depending on whether spin-rotation-symmetry is preserved or not. The negative quantum correction is called \textit{weak localization} and the positive quantum correction is called \textit{weak antilocalization}. An experimental demonstration \cite{Cha94} of the suppression of weak localization by a magnetic field is shown in Fig.\ \ref{fig_Chang}. The measured magnitude $\delta G$ of the peak around zero magnetic field is $0.2\,G_{0}$, somewhat smaller than the fully phase-coherent value of $\frac{1}{4}\,G_{0}$.

\subsection{Sub-Poissonian shot noise}\label{shot_qd}

For $N_{1}=N_{2}\equiv N\gg 1$ the density of transmission eigenvalues for a quantum dot, following from Eq.\ \eqref{PTcircular}, has the form
\begin{equation}
\rho(T)=\frac{N}{\pi}\frac{1}{\sqrt{T}\sqrt{1-T}}.\label{rhoTqd}
\end{equation}
It is different from the result \eqref{rhoresult} for a wire, but it has the same bimodal structure: While the average transmission $\bar{T}=1/2$, the eigenvalue density is peaked at zero and unit transmission.

\begin{figure}[!tb]
\unitlength1cm
\centerline{
\epsfig{figure=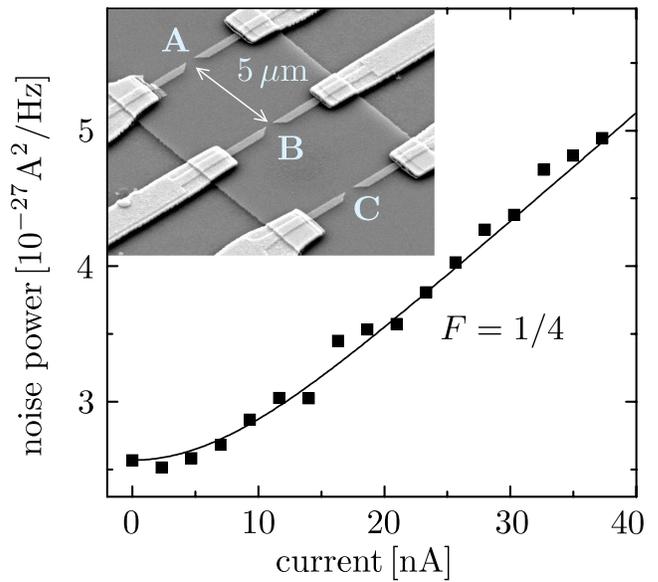,width=20pc}
}
\caption{
Sub-Poissonian shot noise in a quantum dot at $270\,{\rm mK}$. The slope at high currents corresponds to a one-quarter Fano factor, as predicted by RMT. The inset shows an electron micrograph of the device. The quantum dot is contained between point contacts A and B. (The gate labeled C is not operative in this experiment.) Adapted from Ref.\ \cite{Obe01}.
\label{fig_schon}}
\end{figure}

This bimodal structure can be detected as sub-Poissonian shot noise. Instead of Eq.\ \eqref{Fonethird} one now has \cite{Jal94}
\begin{equation}
F\rightarrow 1-\frac{\int dT\,\rho(T)T^{2}}{\int dT\,\rho(T)T}=\frac{1}{4}.\label{Fonequarter}
\end{equation}
An experimental demonstration is shown in Fig.\ \ref{fig_schon}. 

\subsection{Thermopower distribution}\label{thermo}

%reformatted
\begin{figure*}[!tb]
\unitlength1cm
\centerline{
\epsfig{figure=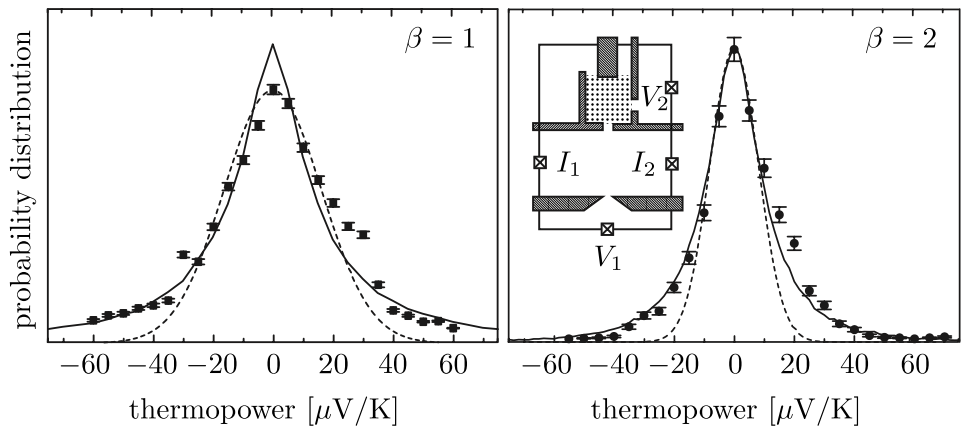,width=30pc}
}
\caption{
Thermopower distribution for $\beta=1$ ($|B|\leq 40\,{\rm mT}$) and $\beta=2$ ($|B|\geq 50\,{\rm mT}$). Experimental results (dots), RMT results (solid line), and Gaussian fit (dashed line) are compared. The inset shows the experimental layout. The crosses denote Ohmic contacts to the two-dimensional electron gas and the shaded areas denote gate electrodes. The heating current is applied between $I_{1}$ and $I_{2}$, while the thermovoltage is 
measured between $V_{1}$ and $V_{2}$. The quantum dot is indicated by the dotted area. Adapted from Ref.\ \cite{God99}.
\label{fig_thermo}}
\end{figure*}

Knowledge of the distribution of the scattering matrix $S(E)$ at a single energy $E=E_{F}$ is sufficient to determine the conductance distribution, but other transport properties require also information on the energy dependence of $S$. The thermopower ${\cal P}$ (giving the voltage produced by a temperature difference at zero electrical current) is a notable example. Since ${\cal P}\propto d\ln G/dE$, we need to know the joint distribution of $S$ and $dS/dE$ at $E_{F}$ to determine the distribution of ${\cal P}$.

This problem goes back to the early days of RMT \cite{Wig55,Smi60}, in connection with the question: What is the time delay experienced by a scattered wave packet? The delay times $\tau_{n}$ are the eigenvalues of the Hermitian matrix product $Q_{\rm WS}=-i\hbar S^{\dagger}dS/dE$, known as the Wigner-Smith matrix in the context of RMT. [For applications in other contexts, see Refs.\ \cite{Das69,Bla91,Got08}.] The solution to the problem of the joint distribution of $S$ and $dS/dE$ (for $S$ in the circular ensemble) was given in Ref.\ \cite{Bro97}. The symmetrized matrix product
\begin{equation}
Q=-i\hbar S^{-1/2}\frac{dS}{dE}S^{-1/2}\label{Qdef}
\end{equation}
has the same eigenvalues as $Q_{\rm WS}$, but unlike $Q_{\rm WS}$ was found to be statistically independent of $S$. The eigenvalues of $Q$ have distribution
%reformatted
\begin{align}
  P(\{\gamma_{n}\}) \propto{}&
  \prod_{i<j} |\gamma_{i} - \gamma_{j}|^{\beta}
  \prod_k \gamma_{k}^{\beta (N_{1}+N_{2})/2}
  e^{-{\beta \tau_H \gamma_k/2}},\nonumber\\
  &\gamma_{n}\equiv 1/\tau_{n}>0. \label{eq:Lag}
\end{align}
The Heisenberg time $\tau_{H}=2\pi\hbar/\delta$ is inversely proportional to the mean level spacing $\delta$ in the quantum dot. Eq.\ \eqref{eq:Lag} is known in RMT as the Laguerre ensemble.

The thermopower distribution following from the Laguerre ensemble is strongly non-Gaussian for small $N_{1}=N_{2}\equiv N$. For $N=1$ it has a cusp at ${\cal P}=0$ when $\beta=1$ and algebraically decaying tails $\propto |{\cal P}|^{-1-\beta}\ln|{\cal P}|$. Significant deviations from a Gaussian are seen in the experiment \cite{God99} shown in Fig.\ \ref{fig_thermo}, for $N=2$.

\subsection{Quantum-to-classical transition}\label{qtocl}

\begin{figure}[!tb]
\unitlength1cm
\centerline{
\epsfig{figure=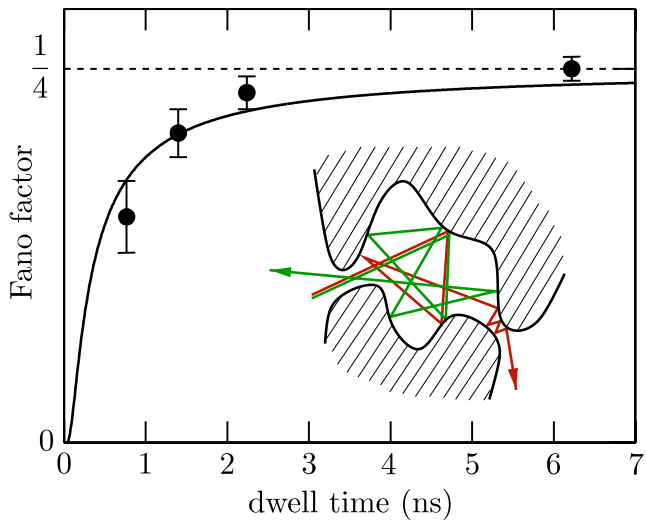,width=20pc}
}
\caption{
Dependence of the Fano factor $F$ of a ballistic chaotic quantum dot on the average
time that an electron dwells inside. The data points
with error bars are measured in a quantum dot, the solid
curve is the theoretical prediction \eqref{Fsuppression} for the quantum-to-classical transition
(with Ehrenfest time $\tau_{E}=0.27\,{\rm ns}$ as a fit parameter). The dwell time \eqref{taudwell} is varied experimentally by changing the number of modes $N$ transmitted through each of the point contacts.
The inset shows graphically the sensitivity to initial conditions of
the chaotic dynamics. Adapted from Ref.\ \cite{Aga00}, with
experimental data from Ref.\ \cite{Obe02}.
\label{fig_AO}}
\end{figure}

RMT is a quantum mechanical theory which breaks down in the classical limit $h\rightarrow 0$. For electrical conduction through a quantum dot, the parameter which governs the quantum-to-classical transition is the ratio $\tau_{E}/\tau_{\rm dwell}$ of Ehrenfest time and dwell time \cite{Aga00}. 

The dwell time $\tau_{\rm dwell}$ is the average time an electron spends inside the quantum dot between entrance and exit through one of the two $N$-mode point contacts. It is given by
\begin{equation}
\tau_{\rm dwell}=\pi\hbar/N\delta.\label{taudwell}
\end{equation} 
The Ehrenfest time $\tau_{E}$ is the time up to which a wave packet follows classical equations of motion, in accord with Ehrenfest's theorem \cite{Ber78,Chi71}. For chaotic dynamics with Lyapunov exponent\footnote{
The Lyapunov exponent $\alpha$ of chaotic motion quantifies the exponential divergence of two trajectories displaced by a distance $\Delta x(t)$ at time $t$, according to $\Delta x(t)=\Delta x(0)e^{\alpha t}$.} 
$\alpha$, it is given by \cite{Sil03}
\begin{equation}
\tau_{E}=\alpha^{-1}\max\bigl[0,\ln(NW/{\cal A}^{1/2})\bigr].\label{tauE}
\end{equation}
Here ${\cal A}$ is the area of the quantum dot and $W$ the width of the $N$-mode point contacts.

The RMT result $F=1/4$ holds if $\tau_{E}\ll\tau_{\rm dwell}$. For longer $\tau_{E}$, the Fano factor is suppressed exponentially \cite{Aga00},
\begin{equation}
F=\tfrac{1}{4}e^{-\tau_{E}/\tau_{\rm dwell}}.\label{Fsuppression}
\end{equation}
This equation expresses the fact that the fraction $1-e^{-\tau_{E}/\tau_{\rm dwell}}$ of electrons that stay inside the quantum dot for times shorter than $\tau_{E}$ follow a deterministic classical motion that does not contribute to the shot noise. RMT applies effectively only to the fraction $e^{-\tau_{E}/\tau_{\rm dwell}}$ of electrons that stay inside for times longer than $\tau_{E}$. The shot noise suppression \eqref{Fsuppression} is plotted in Fig.\ \ref{fig_AO}, together with supporting experimental data \cite{Obe02}.

\sect{Superconductors}\label{Superconductors}

\subsection{Proximity effect}\label{prox}

%reformatted
\begin{figure*}[!tb]
\unitlength1cm
\centerline{
\epsfig{figure=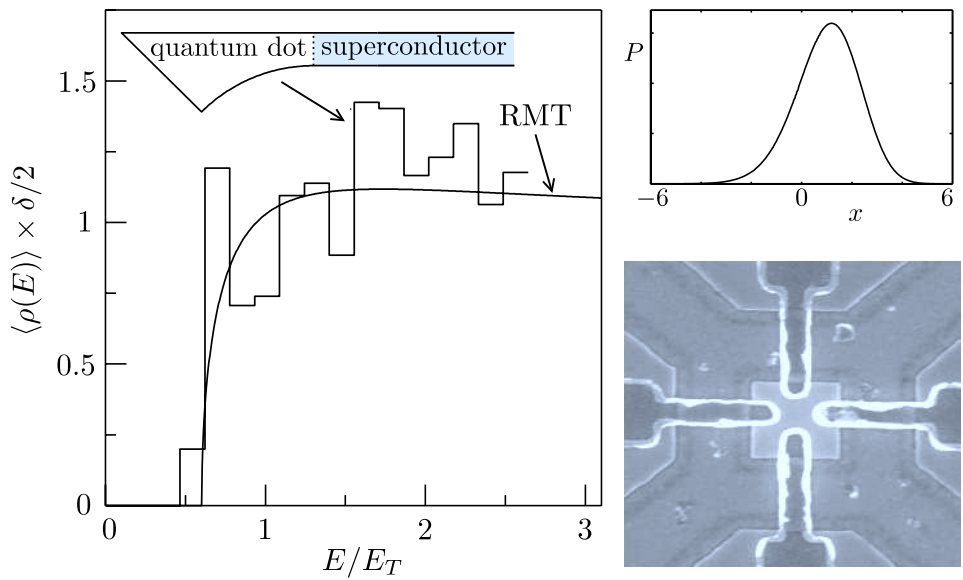,width=30pc}
}
\caption{
\textit{Left panel:} Average density of states (scaled by the Thouless energy $E_{T}=N\delta/4\pi$) of a quantum dot coupled by a ballistic
$N$-mode point contact to a superconductor. The histogram is a numerical calculation for the geometry indicated in the inset (with $N=20$), while the curve is the analytical prediction from RMT. Adapted from Ref.\ \cite{Mel96}. \textit{Upper right panel:} Probability distribution of the lowest excitation energy $E_{1}$, rescaled as $x=(E_{1}-E_{\rm gap})/\Delta_{\rm gap}$. Adapted from Ref.\ \cite{Vav01}. \textit{Lower right panel:} Quantum dot (central square of dimensions $500\,{\rm nm}\times 500\,{\rm nm}$)
fabricated in an InAs/AlSb heterostructure and contacted by four
superconducting Nb electrodes. Device made by A.T. Filip, Groningen University. Figure from Ref.\ \cite{Bee05}.
\label{fig_NS}}
\end{figure*}

Fig.\ \ref{fig_NS} (lower right panel) shows a quantum dot with superconducting electrodes. Without the superconductor the energy spectrum of an ensemble of such quantum dots has GOE statistics.
The proximity of a superconductor has a drastic effect on the energy spectrum, by opening up a gap at the Fermi level. The RMT of this proximity effect was developed in Ref.\ \cite{Mel96} (see Ref.\ \cite{Bee05} for a review). 

A quantum dot coupled to a superconductor has a discrete spectrum for energies below the gap $\Delta$ of the superconductor, given by the roots of the determinantal equation
%reformatted
\begin{align}
&{\rm Det}\,\left[{\mathbb 1}-\alpha(E)^{2}S(E)S(-E)^{\ast}\right]=0,\nonumber\\
&\qquad \alpha(E)=\frac{E}{\Delta}-
i\sqrt{1-\frac{E^{2}}{\Delta^{2}}}.\label{discreteE}
\end{align}
The scattering matrix $S$ (at an energy $E$ measured relative to the Fermi level) describes the coupling of the quantum dot to the superconductor via an $N$-mode point contact and is related to the Hamiltonian $H$ of the isolated quantum dot by Eq.\ \eqref{SHeq}. At low energies $E\ll\Delta$ the energy levels can be obtained as the eigenvalues $E_{i}$ of the effective Hamiltonian
\begin{equation}
H_{\rm eff}=\left(\begin{array}{cc}
H&-\pi WW^{T}\\-\pi W^{\ast}W^{\dagger}&-H^{\ast}
\end{array}\right).\label{calHeffdef}
\end{equation}

The Hermitian matrix $H_{\rm eff}$ is
antisymmetric under the combined operation of charge conjugation (${\cal C}$)
and time inversion (${\cal T}$) \cite{Alt96}:
\begin{equation}
H_{\rm eff}=-\sigma_{y}H_{\rm eff}^{\ast}\sigma_{y},\;\;
\sigma_{y}=
\left(\begin{array}{cc}
0&-i\\
i& 0\\
\end{array}\right).\label{CTsymmetry}
\end{equation}
(An $M\times M$ unit matrix in each of the four blocks of $\sigma_{y}$ is
implicit.) The ${\cal CT}$-antisymmetry ensures that the eigenvalues lie
symmetrically around $E=0$. Only the positive eigenvalues are retained in the
excitation spectrum, but the presence of the negative eigenvalues is felt as a
level repulsion near $E=0$.

As illustrated in Fig.\ \ref{fig_NS} (left panel), the unique feature of the proximity effect is that this level repulsion can extend over energy scales much larger than the mean level spacing $\delta$ in the isolated quantum dot --- at least if time reversal symmetry is not broken. A calculation of the density of states $\langle\rho(E)\rangle=\langle\sum_{i}\delta(E-E_{i})\rangle$ of $H_{\rm eff}$, averaged over $H$ in the GOE, produces a square root singularity in the large-$N$ limit:
\begin{equation}
\langle\rho(E)\rangle\rightarrow\frac{1}{\pi}\sqrt{\frac{E-E_{\rm
gap}}{\Delta^3_{\rm gap}}},\;\;E\rightarrow E_{\rm gap},\;\;N\rightarrow\infty,\label{rhoEneargap}
\end{equation}
If the point contact between quantum dot and superconductor is ballistic ($\Gamma_{n}=1$ for $n=1,2,\ldots N$) the two energies $E_{\rm gap}$ and $\Delta_{\rm gap}$ are given by \cite{Mel96}
\begin{equation}
E_{\rm gap}=\frac{\gamma^{5/2}N\delta}{2\pi}=0.048\,N\delta,\;\;\Delta_{\rm gap}=0.068\,N^{1/3}\delta.\label{EDeltagap}
\end{equation} 
(Here $\gamma = \frac{1}{2}(\sqrt{5} - 1)$ is the golden number.) The gap $E_{\rm gap}$ in the spectrum of the quantum dot is larger than $\delta$ by factor of order $N$.

\subsection{Gap fluctuations}\label{gapfl}

The value \eqref{EDeltagap} of the excitation gap is representative for an ensemble of quantum dots, but each member of the ensemble will have a smallest excitation energy $E_{1}$ that will be slightly different from $E_{\rm gap}$. The distribution of the gap fluctuations is identical upon
rescaling to the known distribution \cite{Tra94} of the lowest eigenvalue in
the GOE \cite{Vav01,Ost01,Lam01}. Rescaling amounts to a change of variables from
$E_{1}$ to $x=(E_{1}-E_{\rm gap})/\Delta_{\rm gap}$, where $E_{\rm gap}$ and
$\Delta_{\rm gap}$ parameterize the square-root dependence \eqref{rhoEneargap}. The probability distribution $P(x)$ of the rescaled gap fluctuations is shown in Fig.\ \ref{fig_NS} (upper right panel).
The gap fluctuations are a mesoscopic, rather than a microscopic effect, because the
typical magnitude $\Delta_{\rm gap}\simeq E_{\rm gap}^{1/3}\delta^{2/3}$ of the
fluctuations is $\gg\delta$ for $E_{\rm gap}\gg \delta$. Still, the
fluctuations are small on the scale of the gap itself.

\subsection{From mesoscopic to microscopic gap}\label{mesotomicro}

The mesoscopic excitation gap of order $N\delta$ induced by the proximity to a superconductor is strongly reduced if time reversal symmetry is broken by application of a magnetic field ($\beta=2$). Because the repulsion of levels at $\pm E$ persists, as demanded by the ${\cal CT}$-antisymmetry \eqref{CTsymmetry}, a microscopic gap around zero energy of order $\delta$ remains. An alternative way to reduce the gap from $N\delta$ to $\delta$, without breaking time reversal symmetry ($\beta=1$), is by contacting the quantum dot to a pair of superconductors with a phase difference of $\pi$ in the order parameter. As shown by Altland and Zirnbauer \cite{Alt96}, the level statistics near the Fermi energy in these two cases is governed by the distribution
\begin{equation}
P(\{E_{n}\}) \propto
  \prod_{i<j} |E^{2}_{i} - E^{2}_{j}|^{\beta}
  \prod_k |E_{k}|^{\beta}
  e^{-c^{2}E_{k}^{2}}, \label{AZ_Lag}
\end{equation}
related to the Laguerre ensemble by a change of variables ($E_{n}^{2}\rightarrow x_{n}$). (The coefficient $c$ is fixed by the mean level spacing in the isolated quantum dot.) The density of states near zero energy vanishes as $|E|^{\beta}$. Two more cases are possible when spin-rotation symmetry is broken, so that in total the three Wigner-Dyson symmetry classes without superconductivity are expanded to four symmetry classes as a consequence of the ${\cal CT}$-antisymmetry.

\subsection{Quantum-to-classical transition}\label{qtoclasgap}

\begin{figure}[!tb]
\unitlength1cm
\centerline{
\epsfig{figure=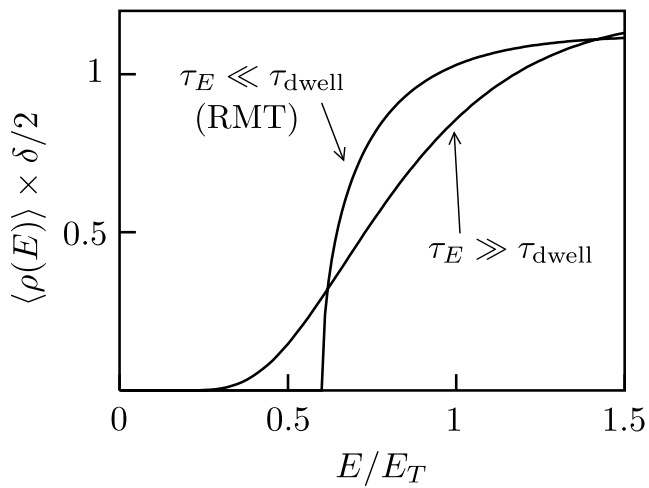,width=20pc}
}
\caption{
Comparison of the density of states \eqref{chaoticdos} with the RMT result \eqref{rhoEneargap}.
These are the two limiting results when the Ehrenfest time
$\tau_{E}$ is, respectively, much larger or much smaller than the mean dwell
time $\tau_{\rm dwell}$.
From Ref.\ \cite{Bee05}.
\label{twoDOS}}
\end{figure}

The RMT of the proximity effect discussed so far breaks down when the dwell time \eqref{taudwell} becomes shorter than the Ehrenfest time \eqref{tauE} \cite{Lod98}. In order of magnitude,\footnote{
More precisely, the gap crosses over between the RMT limit \eqref{EDeltagap} for $\tau_{E}\ll\tau_{\rm dwell}$ and the limit $E_{\rm gap}=\pi\hbar/2\tau_{E}$ for $\tau_{E}\gg\tau_{\rm dwell}$ \cite{Vav03,Bee05,Kui09}.}
 the gap equals $E_{\rm gap}\simeq\min(\hbar/\tau_{E},\hbar/\tau_{\rm dwell})$. In the classical limit $\tau_{E}\rightarrow\infty$, the density of states is given by \cite{Sch99}
\begin{equation}
\langle\rho(E)\rangle=\frac{2}{\delta}\,\frac{(\pi E_{T}/E)^{2}\cosh (\pi
E_{T}/E)}{\sinh^{2} (\pi E_{T}/E)},\label{chaoticdos}
\end{equation}
with $E_{T}=N\delta/4\pi$ the Thouless energy. The density of states \eqref{chaoticdos} (plotted in Fig.\ \ref{twoDOS}) is suppressed exponentially $\propto e^{-\pi E_{T}/E}$ at the Fermi level ($E\rightarrow 0$), but there is no gap.

To understand the absence of a true excitation gap in the limit $\tau_{E}\rightarrow\infty$, we note that in this limit a wave packet follows a classical trajectory in the quantum dot. The duration $t$ of this trajectory, from one reflection at the superconductor to the next, is related to the energy $E$ of the wave packet by $E\simeq\hbar/t$. Since $t$ can become arbitrarily large (albeit with an exponentially small probability $e^{-t/\tau_{\rm dwell}}$), the energy $E$ can become arbitrarily small and there is no gap. 

%%%%%%%%%%%%%%%%%%%%%%%%%%%%%%%%%%%%%%%%%%%%%%%%%%%%%%%%%%%%%%%%%%%%%%%%%%%
\part{Classical and Quantum Optics}
\label{Optics}

\sect{Introduction}\label{intro_optics}

\begin{figure*}[!tb]
\unitlength1cm
\centerline{
\epsfig{figure=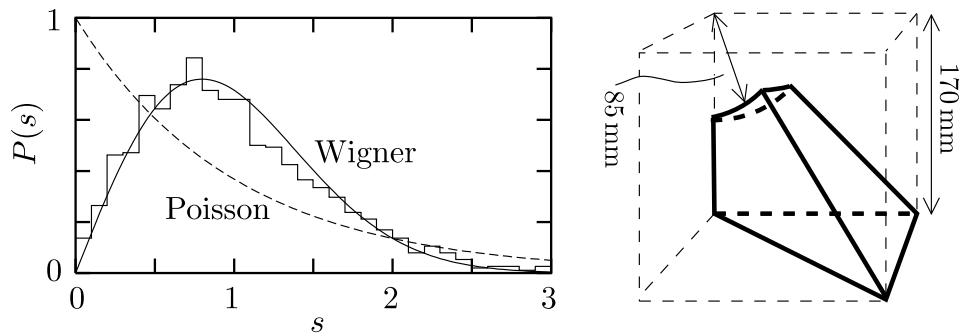,width=30pc}
}
\caption{
Histogram: distribution of spacings $s$ of eigenfrequencies measured in the chaotic microwave resonator shown at the right. (The resonator has superconducting walls, to minimize absorption.) The spacing distribution is close to the Wigner distribution $P(s)\propto s\exp(-\pi s^{2}/4\delta^{2})$ [solid
line] of the GOE, and far from the Poisson distribution $P(s)\propto e^{-s/\delta}$ [dashed
line] of uncorrelated eigenfrequencies. The mean spacing has been set to $\delta=1$, and
non-chaotic ``bouncing-ball'' resonances have been eliminated from the
experimental histogram. Adapted from Ref.\ \cite{Alt97}.
\label{fig_spacing}}
\end{figure*}

Optical applications of random matrix theory came later than electronic applications, perhaps because randomness is much more easily avoided in optics than it is in electronics. The variety of optical systems to which RMT can be applied increased substantially with the realization \cite{Boh84,Ber85} that randomness is not needed at all for GOE statistics of the spectrum. Chaotic dynamics is sufficient, and this is a generic property of resonators formed by a combination of convex and concave surface elements. As an example, we show in Fig.\ \ref{fig_spacing} the Wigner level spacing distribution measured in a microwave cavity with a chaotic shape.

%reformatted

This is an example of an application of RMT to \textit{classical} optics, because the spectral statistics of a cavity is determined by the Maxwell equations of a classical electromagnetic wave. (More applications of this type, including also sound waves, are reviewed in Ref.\ \cite{Kuh05}.) An altogether different type of application of RMT appears in \textit{quantum} optics, when the photon and its quantum statistics play an essential role. Selected applications of RMT to both classical and quantum optics are presented in the following sections. The emphasis is on topics that do not have an immediate analogue in electronics, either because they cannot readily be measured in the solid state or because they involve aspects (such as absorption, amplification or bosonic statistics) that do not apply to electrons.

Some of the concepts used in this Chapter were introduced in the previous Chapter on applications of RMT to condensed matter physics, in particular in Secs.\ \ref{nonlog}, \ref{noise}, and \ref{Gausstocirc}.

\sect{Classical optics}\label{class}

\subsection{Optical speckle and coherent backscattering}\label{coh_back}

%reformatted
\begin{figure*}[!tb]
\unitlength1cm
\centerline{
\epsfig{figure=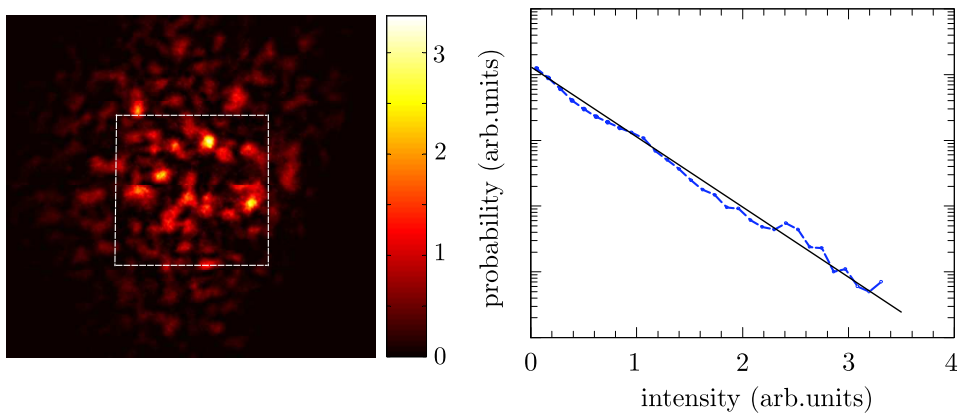,width=30pc}
}
\caption{
\textit{Left panel:} Speckle pattern produced by a laser beam behind a diffusor (full scale $45\,{\rm mrad}\times 45\,{\rm mrad}$). The vertical bar indicates the color coding of the intensity, in arbitrary units. The average angular opening angle $\delta \alpha \approx 1.3\,{\rm mrad}$ of a bright or dark spot (a ``speckle'') is equal to $\lambda/\pi R$, with $\lambda=830\,{\rm nm}$ the wave length and $R= 200\,\mu{\rm m}$ the radius of the illuminated area on the diffusor. The envelope of the intensity pattern reflects the $18\,{\rm mrad}$ opening angle of the directional scattering from this type of diffusor. The intensity distribution $P(I)$ of the speckle pattern measured inside the white square is plotted in the right panel, and compared with the exponential distribution \eqref{PIspeckle} (straight line in the semi-logarithmic plot).
Figure courtesy of M.P. van Exter.
\label{fig_speckle}}
\end{figure*}

Optical speckle, shown in Fig.\ \ref{fig_speckle}, is the random interference pattern that is observed when coherent radiation is transmitted or reflected by a random medium. It has been much studied since the discovery of the laser, because the speckle pattern carries information both on the coherence properties of the radiation and on microscopic details of the scattering object \cite{Goo07}. The superposition of partial waves with randomly varying phase and amplitude produces a wide distribution $P(I)$ of intensities $I$ around the average $\bar{I}$. For full coherence and complete randomization the distribution has the exponential form
\begin{equation}
P(I)=\bar{I}^{-1}\exp(-I/\bar{I}),\;\;I>0.\label{PIspeckle}
\end{equation}

For a description of speckle in the framework of RMT \cite{Mel88a}, it is convenient to enclose the scattering medium in a wave guide containing a large number $N$ of propagating modes. The reflection matrix $r$ is then an $N\times N$ matrix with random elements. Time-reversal symmetry (reciprocity) dictates that $r$ is symmetric. Deviations of $r$ from unitarity can be ignored if the mean free path $l$ is much smaller than both the length $L$ of the scattering medium and the absorption length $l_{a}$. The RMT assumption is that $r$ is distributed according to the circular orthogonal ensemble (COE), which means that $r=UU^{T}$ with $U$ uniformly distributed in the group ${\cal U}(N)$ of $N\times N$ unitary matrices.

\begin{figure}[!tb]
\unitlength1cm
\centerline{
\epsfig{figure=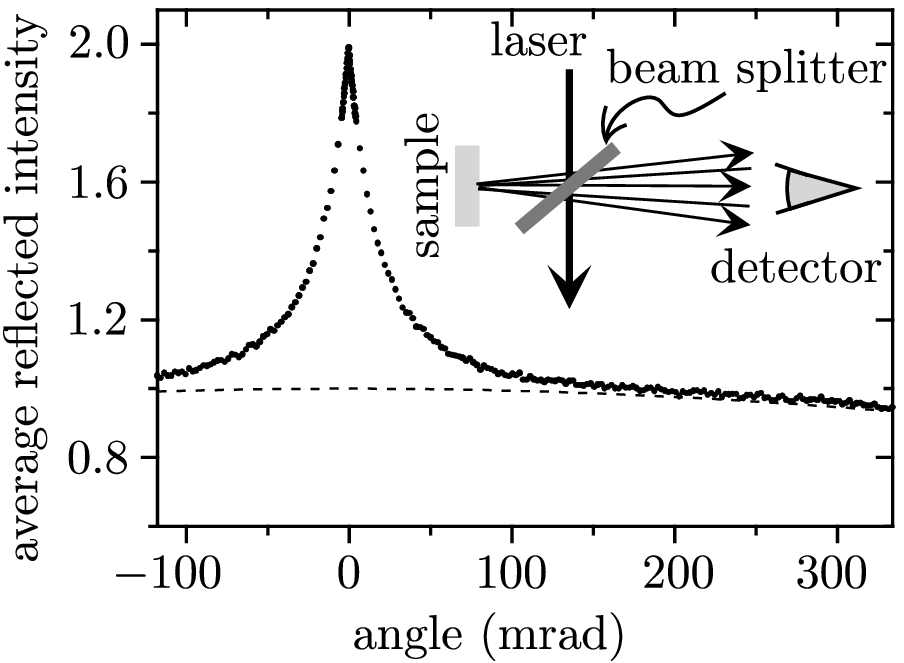,width=20pc}
}
\caption{
Measurement of coherent backscattering from a ZnO powder. The sample is rotated to average the reflected intensity, which is plotted against the scattering angle. The measured peak due to coherent backscattering is superimposed on the diffuse scattering intensity (dashed curve, normalized to unity in the backscattering direction at zero angle). The relative height of the peak is a factor-of-two, in accord with Eq.\ \eqref{Iaverage}. The angular width is of order $1/kl\approx 40\,{\rm mrad}$, for wave length $\lambda=2\pi/k=514\,{\rm nm}$ and mean free path $l=1.89\,\mu{\rm m}$. The inset shows the optical setup. Use of a beam splitter permits detection in the backscattering direction, which would otherwise be blocked by the incident laser beam. Adapted from Ref.\ \cite{Wie95}.
\label{fig_backsc}}
\end{figure}

In this description, the reflected intensity in mode $n$ for a wave incident in mode $m$ is given by $I_{nm}=|r_{nm}|^{2}$. The intensity distribution can be easily calculated in the limit $N\rightarrow\infty$, when the complex matrix elements $r_{nm}$ with $n\leq m$ have independent Gaussian distributions of zero mean and variance\footnote{
For an introduction to such integrals over the unitary group, see Ref.\ \cite{Bee97}. The factor $N+1$ in the denominator ensures that $\sum_{m=1}^{N}|r_{nm}|^{2}=1$, as required by unitarity, but the difference between $N$ and $N+1$ can be neglected in the large-$N$ limit.}
\begin{equation}
\langle |r_{nm}|^{2}\rangle=\int_{{\cal U}(N)}\!\!dU\, \sum_{k,k'=1}^{N}U_{nk}^{\vphantom{\ast}}U_{mk'}^{\vphantom{\ast}}U^{\ast}_{nk'}U^{\ast}_{mk}=\frac{1+\delta_{nm}}{N+1}.\label{Iaverage}
\end{equation}
The resulting distribution of $I_{nm}$ in the large-$N$ limit has the exponential form \eqref{PIspeckle}, with an average intensity $\bar{I}_{nm}=(1+\delta_{nm})N^{-1}$ which is twice as large when $n=m$ than when $n\neq m$. This doubling of the average reflected intensity at the angle of incidence is the \textit{coherent backscattering} effect \cite{Akk07}, illustrated in Fig.\ \ref{fig_backsc}.

The RMT assumption of a COE distribution of the reflection matrix correctly reproduces the height of the coherent backscattering peak, but it cannot reproduce its width \cite{Akk88,Mel88b}. The Kronecker delta in Eq.\ \eqref{Iaverage} would imply an angular opening $\delta\alpha\simeq 1/kW$ of the peak (for light of wave number $k$ in a wave guide of width $W$). This is only correct if the mean free path $l$ is larger than $W$. In a typical experiment $l\ll W$ and the angular opening is $\delta\alpha\simeq 1/kl$ (as it is in Fig.\ \ref{fig_backsc}).

\subsection{Reflection from an absorbing random medium}\label{optically_active}

An absorbing medium has a dielectric constant $\varepsilon$ with a positive imaginary part. The intensity of radiation which has propagated without scattering over a distance $L$ is then multiplied by a factor $e^{-\sigma L}$. The decay rate $\sigma>0$ at wave number $k$ is related to the dielectric constant by $\sigma=2k\,{\rm Im}\,\sqrt{\varepsilon}$.

The absence of a conservation law in an absorbing medium breaks the unitarity of the scattering matrix. The circular orthogonal ensemble, of uniformly distributed symmetric unitary matrices, should therefore be replaced by another ensemble. The appropriate ensemble was derived in Refs.\ \cite{Bee96,Bru96}, for the case of reflection from an infinitely long absorbing wave guide. The result is that the $N$ eigenvalues $R_{n}\in[0,1]$ of the reflection matrix product $rr^{\dagger}$ are distributed according to the Laguerre orthogonal ensemble, after a change of variables to $\lambda_{n}=R_{n}(1-R_{n})^{-1}\geq 0$:
\begin{equation}
P(\{\lambda_{n}\})\propto\prod_{i<j}|\lambda_{j}-\lambda_{i}|
\prod_{k}\exp[-\sigma l(N+1)\lambda_{k}].\label{Pwire}
\end{equation}

The distribution \eqref{Pwire} is obtained by including an absorption term into the DMPK equation \eqref{FPxa}. This loss-drift-diffusion equation has the form \cite{Bee96,Bru96}
%reformatted
\begin{align}
l\frac{\partial P}{\partial L}=&{}\frac{2}{\beta N+2-\beta}\sum_{n=1}^{N}
\frac{\partial}{\partial\lambda_{n}}\lambda_{n}(1+\lambda_{n})\nonumber\\
&\times\Bigl[J\frac{\partial}{\partial\lambda_{n}}\frac{P}{J}+\sigma l(\beta N+2-\beta)P\Bigr],\nonumber\\
&{\rm with}\;\;J=\prod_{i<j}|\lambda_{j}-\lambda_{i}|^{\beta}.\label{FP}
\end{align}
The drift-diffusion equation \eqref{FPxa} considered in the electronic context is obtained by setting $\sigma=0$ and transforming to the variables $x_{n}=\sinh^{2}\lambda_{n}$.

In the limit $L\rightarrow\infty$ we may equate the left-hand-side of Eq.\ \eqref{FP} to zero, and we arrive at the solution \eqref{Pwire} for $\beta=1$ (unbroken time reversal symmetry). More generally, for any $\beta$, the distribution of the $R_{n}$'s in the limit $L\rightarrow\infty$ can be written in the form of a Gibbs distribution at a fictitious temperature $\beta^{-1}$,
%reformatted
\begin{align}
P(\{R_{n}\})\propto{}&\exp\Bigl[-\beta\Bigl(\sum_{i<j}
u(R_{i},R_{j})+\sum_{i}V(R_{i})\Bigr)\Bigr],
\label{PRglobala}\\
u(R,R')={}&-\ln|R-R'|,\nonumber\\
V(R)={}&\left(N-1+\frac{2}{\beta}\right)\left[\frac{\sigma lR}{1-R}+\ln(1-R)\right].
\label{PRglobalb}
\end{align}
The eigenvalue interaction potential $u(R,R')$ is \textit{logarithmic}. This can be contrasted with the \textit{nonlogarithmic} interaction potential in the absence of absorption, discussed in Sec.\ \ref{nonlog}. Because $R_{n}=1-T_{n}$ without absorption, the interaction potential \eqref{final3u} of that section can be written as
\begin{equation}
u(R,R')=-{\textstyle\frac{1}{2}}
\ln|R-R'|
-{\textstyle\frac{1}{2}}\ln|x-x'|,\;\;{\rm with}\;\;R\equiv \tanh^{2}x.\label{final3uR}
\end{equation}

As calculated in Ref.\ \cite{Mis96}, the change in interaction potential has an observable consequence in the sample-to-sample fluctuations of the reflectance
\begin{equation}
{\cal R}={\rm Tr}\,rr^{\dagger}=\sum_{n=1}^{N}R_{n}.\label{calRdef}
\end{equation}
With increasing length $L$ of the absorbing disordered waveguide, the variance of the reflectance drops from the value ${\rm Var}\,{\cal R}=2/15\beta$ associated with the nonlogarithmic interaction \eqref{final3uR} [cf.\ Eq.\ \eqref{UCF}], to the value ${\rm Var}\,{\cal R}=1/8\beta$ for a logarithmic interaction [cf.\ Eq.\ \eqref{UCFglobal}]. The crossover occurs when $L$ becomes longer than the absorption length $l_{a}=\sqrt{l/\sigma_{a}}$, in the large-$N$ regime $N\gg 1/\sqrt{\sigma l}\gg 1$.

\subsection{Long-range wave function correlations}\label{longrange}

\begin{figure}[!tb]
\unitlength1cm
\centerline{
\epsfig{figure=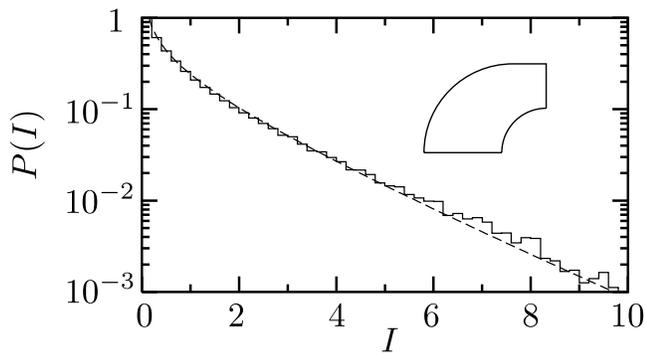,width=20pc}
}
\caption{
Comparison of the Porter-Thomas
distribution \eqref{PorterThomas} [dashed line] of wave function intensities $I$ in the GOE, with the intensity distribution measured on the two-dimensional microwave cavity shown in the
inset. (The average intensity has been set to $\bar{I}=1$.) Adapted from Ref.\ \cite{Kud95}.
\label{fig_porter}}
\end{figure}

The statistics of wave function intensities $I=|\Psi(\bm{r})|^{2}$ in a chaotic cavity is described by the Porter-Thomas distribution \cite{Por65},
\begin{equation}
P(I)=(2\pi \bar{I})^{-1/2}I^{-1/2}\exp(-I/2\bar{I}),\;\;I>0,\label{PorterThomas}
\end{equation}
with $\bar{I}$ the average intensity. Eq.\ \eqref{PorterThomas} assumes time reversal symmetry, so $\Psi$ is real (symmetry index $\beta=1$). An experimental demonstration in a microwave resonator is shown in Fig.\ \ref{fig_porter}. 

In the context of RMT, the distribution \eqref{PorterThomas} follows from the GOE ensemble of the real symmetric $M\times M$ matrix $H$ (the effective Hamiltonian), which determines the eigenstates of the cavity. The intensity $I$ corresponds to the square of a matrix element $O_{nm}$ of the orthogonal matrix which diagonalizes $H$, where the index $n$ labels a point in discretized space and the index $m$ labels a particular eigenstate. In the large-$M$ limit the matrix elements of $O$ have a Gaussian distribution, which implies Eq.\ \eqref{PorterThomas} for the distribution of $I=O_{nm}^{2}$. 

Different matrix elements $O_{nm}$ and $O_{n'm}$ are independent, so the wave function has no spatial correlations in the RMT description. This is an approximation, but since the actual correlations decay on the scale of the wave length \cite{Ber77}, it is accurate to say that there are no \textit{long-range} wave function correlations in a chaotic cavity.

The same absence of long-range correlations applies if time reversal symmetry is fully broken, by the introduction of a sufficiently strong magneto-optical element in the cavity \cite{Sto99}. The intensity distribution changes from the Porter-Thomas distribution \eqref{PorterThomas} to the exponential distribution \eqref{PIspeckle}, but spatial correlations still decay on the scale of the wave length. \textit{Partially} broken time reversal symmetry, however, has the striking effect of introducing wave function correlations that persist throughout the entire cavity. This was discovered theoretically by Fal'ko and Efetov \cite{Fal94} for the crossover from GOE to GUE.

%reformatted
\begin{figure*}[!tb]
\unitlength1cm
\centerline{
\epsfig{figure=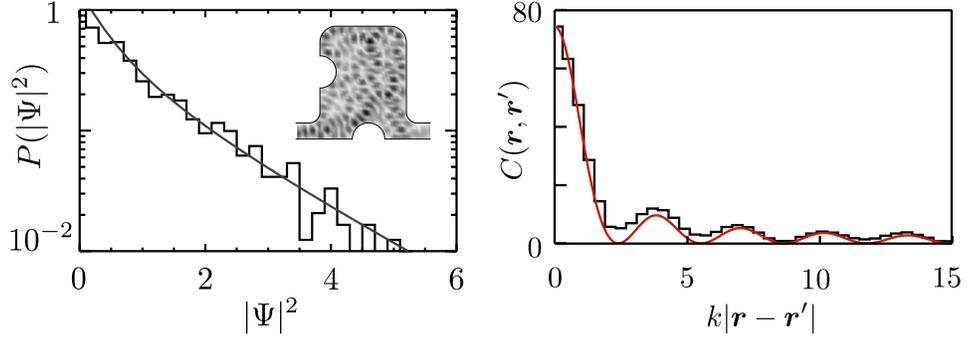,width=30pc}
}
\caption{
\textit{Left panel:} Distribution of the intensity $|\Psi(\bm{r})|^{2}$ of a traveling wave at a fixed frequency in the open two-dimensional chaotic microwave cavity shown in the inset (dimensions $21\,{\rm cm}\times 18\,{\rm cm}$). The wave function $\Psi$ is the component of the electric field perpendicular to the cavity (normalized to unit average intensity), for a wave traveling from the right to the left lead. The measured values (histogram) are compared with the distribution \eqref{PIrho} (solid curve), fitted to a phase rigidity $|\rho|^{2}=0.5202$. The grey scale plot in the inset shows the spatial intensity variations, with black corresponding to maximal intensity.
\textit{Right panel:} Correlator of squared intensity, for a single mode in both the right and left leads. The histogram shows the measured correlator, averaged over position in the cavity and frequency of the traveling wave. The solid curve is the theoretical prediction \cite{Bro03}, which tends to the nonzero limit $0.078$ for $k|\bm{r}-\bm{r}'|\gg 1$.
Adapted from Ref.\ \cite{Kim05}.
\label{fig_long_range}}
\end{figure*}

An altogether different way to partially break time reversal symmetry is to open up the cavity by attaching a pair of $N$-mode leads to it, and to excite a traveling wave from one lead to the other \cite{Pni96}. Brouwer \cite{Bro03} found that, if $N$ is of order unity, the traveling wave produces relatively large long-range wave function correlations inside the cavity. As shown in Fig.\ \ref{fig_long_range}, these correlations have been measured in a microwave resonator \cite{Kim05}.

Partially broken time reversal symmetry means that a wave function $\Psi(\bm{r})$ is neither real nor fully complex. Following Ref.\ \cite{Lan97}, the crossover from real to fully complex wave functions is quantified by the phase rigidity
\begin{equation}
\rho=\frac{\int d\bm{r}\,\Psi(\bm{r})^{2}}{\int d\bm{r}\,|\Psi(\bm{r})|^{2}}.\label{rhodef}
\end{equation}
A real wave function has $\rho=1$ while a fully complex wave function has $\rho=0$. 

As $|\rho|$ decreases from 1 to 0, the intensity distribution crosses over from the Porter-Thomas distribution \eqref{PorterThomas} to the exponential distribution \eqref{PIspeckle}, according to \cite{Pni96}
\begin{equation}
P(I|\rho)=\frac{1}{\bar{I}\sqrt{1-|\rho|^{2}}}\exp\left(-\frac{I/\bar{I}}{1-|\rho|^{2}}\right)I_{0}\left(\frac{|\rho|I/\bar{I}}{1-|\rho|^{2}}\right).\label{PIrho}
\end{equation}
(The function $I_{0}$ is a Bessel function.) The notation $P(I|\rho)$ indicates that this is the intensity distribution for an eigenstate with a given value of $\rho$. The distribution $P(\rho)$ of $\rho$ among different eigenstates, calculated in Ref.\ \cite{Bro03}, is broad for $N$ of order unity. 

For any given phase rigidity the joint distribution of the intensities $I\equiv I(\bm{r})$ and $I'\equiv I(\bm{r}')$ factorizes if $k|\bm{r}-\bm{r}'|\gg 1$. The long-range correlations appear upon averaging over the broad distribution of phase rigidities, since
\begin{equation}
P(I,I')=\int d\rho\,P(\rho)P(I|\rho)P(I'|\rho)\label{PII}
\end{equation}
no longer factorizes.

\subsection{Open transmission channels}\label{openchannels}

The bimodal transmission distribution \eqref{rhoresult}, first obtained by Dorokhov in 1984 \cite{Dor84}, tells us that a fraction $l/L$ of the transmission eigenvalues through a random medium is of order unity, the remainder being exponentially small. A physical consequence of these open channels, discussed in Sec.\ \ref{noise}, is the sub-Poissonian shot noise of electrical current \cite{Bee92}. As expressed by Eq. \eqref{Fonethird}, the shot noise power is reduced by a factor $1-2/3=1/3$, because the spectral average $\overline{T^{2}}$ of  the transmission eigenvalues is $2/3$ of the average transmission $\overline{T}=l/L$. If all transmission eigenvalues would have been close to their average, one would have found $\overline{T^{2}}/\overline{T}\simeq l/L\ll 1$ and the shot noise would have been Poissonian.

The observation of sub-Poissonian shot noise is evidence for the existence of open transmission channels, but it is indirect evidence --- because a theory is required to interpret the observed shot noise in terms of the transmission eigenvalues. In fact, one can alternatively interpret the sub-Poissonian shot noise in terms of a semiclassical theory that makes no reference at all to the transmission matrix \cite{Nag92}. 

%reformatted
\begin{figure*}[!tb]
\unitlength1cm
\centerline{
\epsfig{figure=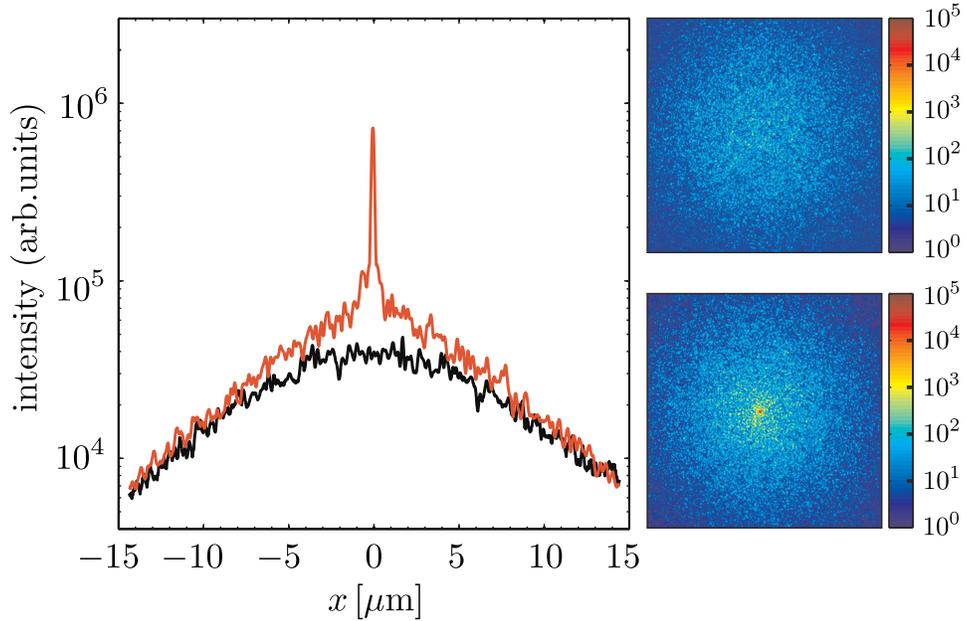,width=30pc}
}
\caption{
\textit{Right panels:} Speckle pattern (area $30\,\mu{\rm m}\times 30\,\mu{\rm m}$) behind a diffusor (a $11.3\,\mu{\rm m}$ layer of ZnO particles with mean free path $l=0.85\,\mu{\rm m}$), for a random incident wave front (top) and for a wave front optimized to couple to open transmission channels (bottom). The intensity of the bright speckle at the center in the bottom panel is a factor of $750$ greater than the background. \textit{Left panel:} Intensity profile, integrated over the $y$-direction to average out the speckle pattern. The optimized wave front (red) has a peak, which the random wave front (black) lacks.
Adapted from Ref.\ \cite{Vel08}.
\label{fig_openchannel}}
\end{figure*}

A direct measurement of the ratio $\overline{T^{2}}/\overline{T}$ would require the preparation of a specific scattering state, which is not feasible in electronics. In optics, however, this is a feasible experiment --- as demonstrated very recently by Vellekoop and Mosk \cite{Vel08}. By adjusting the relative amplitude and phase of a superposition of plane waves, they produced an incident wave with amplitude $E^{\rm in}_{n}=t^{\ast}_{m_{0}n}$ in mode $n=1,2,\ldots N$ (for $N\simeq 10^{4}$). The index $m_{0}$ corresponds to an arbitrarily chosen ``target speckle'' behind a diffusor, located at the center of the square speckle pattern in Fig.\ \ref{fig_openchannel}. The transmitted wave has amplitude
\begin{equation}
E^{\rm out}_{m}=\sum_{n}t_{mn}E^{\rm in}_{n}=(tt^{\dagger})_{mm_{0}}.\label{Eout}
\end{equation}
As shown in Ref. \cite{Vel08}, this optimized incident wave front can be constructed ``by trial and error'' without prior knowledge of the transmission matrix, because it maximizes the transmitted intensity at the target speckle (for a fixed incident intensity). The optimal increase in intensity is a factor of order $Nl/L\simeq 10^{3}$, as observed.

The total transmitted intensity is
\begin{equation}
I^{\rm out}=\sum_{m}|E^{\rm out}_{m}|^{2}=(tt^{\dagger}tt^{\dagger})_{m_{0}m_{0}}.\label{Iout}
\end{equation}
The average transmitted intensity, averaged over the target speckle, gives the spectral average $\overline{T^{2}}$,
\begin{equation}
\overline{I^{\rm out}}=\frac{1}{N}\sum_{m_{0}}I^{\rm out}=\frac{1}{N}{\rm Tr}\,(tt^{\dagger})^{2}=\overline{T^{2}}.\label{Ibarout}
\end{equation}
The average incident intensity is simply $\overline{I^{\rm in}}=N^{-1}{\rm Tr}\,tt^{\dagger}=\overline{T}$, so the ratio of transmitted and incident intensities gives the required ratio of spectral averages, $\overline{I^{\rm out}}/\overline{I^{\rm in}}=\overline{T^{2}}/\overline{T}$. The experimental results are consistent with the value $2/3$ for this ratio, in accord with the bimodal transmission distribution \eqref{rhoresult}.

\sect{Quantum optics}\label{quantum}

\subsection{Grey-body radiation}\label{grey}

The emission of photons by matter in thermal equilibrium is not a series of
independent events. The textbook example is black-body radiation
\cite{Man95}: Consider a system in thermal equilibrium (temperature $T$)
that fully absorbs any incident radiation in $N$ propagating modes
within a frequency interval $\delta\omega$ around $\omega$. A photo\-detector
counts the emission of $n$ photons in this frequency interval during a long
time $t\gg 1/\delta\omega$. The probability distribution $P(n)$ is given by the
negative-binomial distribution with $\nu=Nt\delta\omega/2\pi$ degrees of
freedom,
\begin{equation}
P(n)\propto{n+\nu-1\choose n}\exp(-n\hbar\omega/k_{\rm B}T).\label{binomial}
\end{equation}

The binomial coefficient counts the number of partitions of $n$ bosons among
$\nu$ states. The mean photo\-count $\bar{n}=\nu f$ is proportional to the
Bose-Einstein function
\begin{equation}
f(\omega,T)=[\exp(\hbar\omega/k_{\rm B}T)-1]^{-1}.\label{BEfunction}
\end{equation}
In the limit $\bar{n}/\nu\rightarrow 0$, Eq.\ (\ref{binomial}) approaches the
Poisson distribution $P(n)\propto\bar{n}^{n}/n!$ of independent photo\-counts.
The Poisson distribution has variance ${\rm Var}\,n=\bar{n}$ equal to its mean.
The negative-binomial distribution describes photo\-counts that occur in
``bunches'', leading to an increase of the variance by a factor
$1+\bar{n}/\nu$.

By definition, a black body has scattering matrix $S=0$, because all incident radiation is absorbed. If the absorption is not strong enough, some radiation will be transmitted or reflected and $S$ will differ from zero. Such a ``grey body'' can still be in thermal equilibrium, but the statistics of the photons which its emits will differ from the negative-binomial distribution \eqref{binomial}. A general expression for the photon statistics of grey-body radiation in terms of the scattering matrix was derived in Ref.\ \cite{Bee98}. The expression is simplest in terms of the generating function
\begin{equation}
F(\xi)=\ln\sum_{n=0}^{\infty}(1+\xi)^{n}P(n),\label{Fxidef}
\end{equation}
from which $P(n)$ can be reconstructed via
\begin{equation}
P(n)=\lim_{\xi\rightarrow -1}\frac{1}{n!}\frac{d^{n}}{d\xi^{n}}e^{F(\xi)}.
\end{equation}
The relation between $F(\xi)$ and $S$ is
\begin{equation}
F(\xi)=-\frac{t\delta\omega}{2\pi}
\ln{\rm Det}\,\bigl[{\mathbb 1}-({\mathbb 1}-SS^{\dagger})\xi f\bigr].\label{Fxilong}
\end{equation}

If the grey body is a chaotic resonator, RMT can be used to determine the sample-to-sample statistics of $S$ and thus of the photocount distribution. What is needed is the distribution of the socalled ``scattering strengths'' $\sigma_{1},\sigma_{2},\ldots \sigma_{N}$, which are the eigenvalues of the matrix product $SS^{\dagger}$. All $\sigma_{n}$'s are equal to zero for a black body and equal to unity in the absence of absorption. The distribution function $P(\{\sigma_{n}\})$ is known exactly for weak absorption (Laguerre orthogonal ensemble) and for a few small values of $N$ \cite{Bee01}. In the large-$N$ limit, the eigenvalue density $\rho(\sigma)=\langle\sum_{n}\delta(\sigma-\sigma_{n})\rangle$ is known in closed-form \cite{Bee99}, which makes it possible to compute the ensemble average of arbitrary moments of $P(n)$.

The first two moments are given by
\begin{equation}
\bar{n}=\nu f\frac{1}{N}\sum_{n=1}^{N}(1-\sigma_{n}),\;\;{\rm Var}\,n=\bar{n}+\nu f^{2}\frac{1}{N}\sum_{n=1}^{N}(1-\sigma_{n})^{2}.\label{firsttwomoments}
\end{equation}
For comparison
with black-body radiation we parameterize the variance in terms of the
effective number $\nu_{\rm eff}$ of degrees of freedom \cite{Man95},
\begin{equation}
{\rm Var}\,n=\bar{n}(1+\bar{n}/\nu_{\rm eff}),\label{nueffdef}
\end{equation}
with $\nu_{\rm eff}=\nu$ for a black body. Eq.\ \eqref{firsttwomoments} implies a \textit{reduced} number of degrees of freedom for grey-body radiation,
\begin{equation}
\frac{\nu_{\rm eff}}{\nu}=\frac{\bigl[\sum_{n}(1-\sigma_{n})\bigr]^{2}}{N\sum_{n}
(1-\sigma_{n})^{2}}\leq 1.\label{nueffrho}
\end{equation}
Note that the reduction occurs only for $N>1$.

\begin{figure}[!tb]
\unitlength1cm
\centerline{
\epsfig{figure=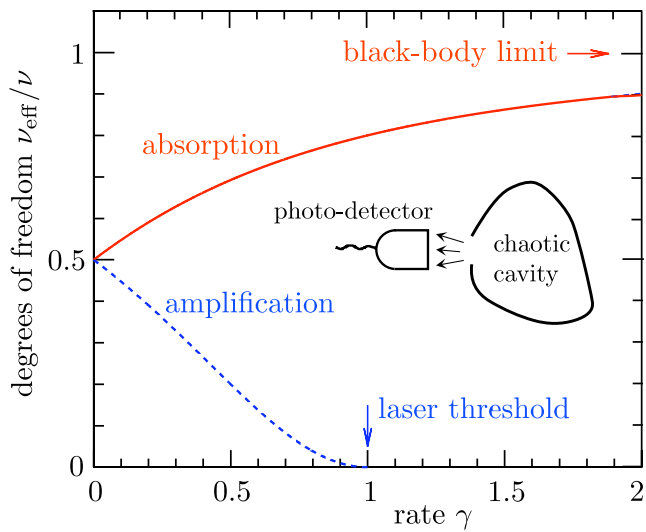,width=20pc}
}
\caption{
Effective number of degrees of freedom as a function of normalized absorption
or amplification rate in a chaotic cavity (inset). The black-body
limit for absorbing systems (red, solid line) and the laser threshold for amplifying systems (blue, dashed line) are
indicated by arrows.
Adapted from Ref.\ \cite{Bee98}.
\label{fig_nueff}}
\end{figure}

The ensemble average for $N\gg 1$ is
\begin{equation}
\nu_{\rm eff}/\nu=(1+\gamma)^{2}(\gamma^{2}+2\gamma+2)^{-1},
\label{nueffcavity}
\end{equation}
with $\gamma=\sigma\tau_{\rm dwell}$ the product of the absorption rate $\sigma$ and the
mean dwell time $\tau_{\rm dwell}\equiv
2\pi/N\delta$ of a photon
in the cavity in the absence of absorption. (The cavity has a mean spacing $\delta$ of eigenfrequencies.) As shown in Fig.\ \ref{fig_nueff} (red solid curve), weak absorption reduces $\nu_{\rm eff}$ by up to a factor of two relative to the black-body value.

So far we have discussed thermal emission from absorbing systems.
The general formula \eqref{Fxilong} can also be applied to amplified spontaneous
emission, produced by a population inversion of the atomic levels in the cavity. The factor $f$ now describes the degree of population inversion of a two-level system, with $f=-1$ for complete inversion (empty lower level, filled upper level). The scattering strengths $\sigma_{n}$ for an amplifying system are $>1$, and in fact one can show that $\sigma_{n}\mapsto 1/\sigma_{n}$ upon changing $\sigma\mapsto -\sigma$ (absorption rate $\mapsto$ amplification rate). As a consequence, Eq.\ \eqref{nueffcavity} can also be applied to an amplifying cavity, if we change $\gamma\mapsto -\gamma$. The result (blue dashed curve in Fig.\ \ref{fig_nueff}) is that the ratio $\nu_{\rm eff}/\nu$ decreases with increasing $\gamma=|\sigma|\tau_{\rm dwell}$ --- vanishing at $\gamma=1$. This is the laser threshold, which we discuss next.

\subsection{RMT of a chaotic laser cavity}\label{lasercavity}

Causality requires that the scattering matrix $S(\omega)$ has all its poles $\Omega_{m}-i\Gamma_{m}/2$ in the lower half of the complex frequency plane. Amplification with rate $\sigma>0$ adds a term $i\sigma/2$ to the poles, shifting them upwards towards the real axis. The laser threshold is reached when the decay rate $\Gamma_{0}$ of the pole closest to the real axis (the ``lasing mode'') equals the amplification rate $\sigma$. For $\sigma>\Gamma_{0}$ the loss of radiation from the cavity is less than the gain due to stimulated emission, so the cavity will emit radiation in a narrow frequency band width around the lasing mode. If the cavity has chaotic dynamics, the ensemble averaged properties of the laser can be described by RMT.\footnote{
The statistical properties of a chaotic laser cavity are closely related to those of socalled random lasers (see Ref.\ \cite{Cao05} for a review of experiments and Ref.\ \cite{Tur08} for a recent theory). The confinement in a random laser is not produced by a cavity, but presumably by disorder and the resulting wave localization. (Alternative mechanisms are reviewed in Ref.\ \cite{Zai09}.)
}

For this purpose, we include amplification in the Weidenm\"{u}ller formula \eqref{SHeq}, which takes the form
\begin{equation}
S(\omega)={\mathbb 1}-2\pi iW^{\dagger}(\omega-i\sigma/2-{\cal H})^{-1}W.\label{ScalH}
\end{equation}
The poles of the scattering matrix are the complex eigenvalues of the $M\times M$ matrix
\begin{equation}
{\cal H}=H-i\pi WW^{\dagger}=U\,{\rm diag}\,(\Omega_{1}-i\Gamma_{1},\ldots,\Omega_{M}-i\Gamma_{M})U^{-1},\label{calHdef}
\end{equation}
constructed from the Hamiltonian $H$ of the closed cavity and the $M\times N$ coupling matrix $W$ to the outside. Because ${\cal H}$ is not Hermitian, the matrix $U$ which diagonalizes ${\cal H}$ is not unitary. In the RMT description one takes a Gaussian ensemble for $H$ and a non-random $W$, and seeks the distribution of eigenvalues and eigenvectors of ${\cal H}$. This is a difficult problem, but most of the results needed for the application to a laser are known \cite{Fyo03}. 

The first question to ask, is at which frequencies the laser will radiate. There can be more than a single lasing mode, when more than a single pole has crossed the real axis. The statistics of the laser frequencies has been studied in Refs.\ \cite{Mis98,Hac05,Zai06}. Only a subset $N_{\rm lasing}$ of the $N_{\sigma}$ modes with $\Gamma_{m}<\sigma$ becomes a laser mode, because of mode competition: If two modes have an appreciable spatial overlap, the mode which starts lasing first will deplete the population inversion before the second mode has a chance to be amplified. For weak coupling of the modes to the outside, when the wave functions have the Porter-Thomas distribution, the average number of lasing modes scales as $\bar{N}_{\rm lasing}\propto\bar{N}_{\sigma}^{2/3}$ \cite{Mis98}.

Once we know the frequency of a lasing mode, we would like to know its width. The radiation from a laser is characterized by a very narrow line width, limited by the vacuum fluctuations of the electromagnetic field. The quantum-limited linewidth, or Schawlow-Townes linewidth \cite{Sch58},
\begin{equation}
\delta\omega= \tfrac{1}{2}K\Gamma_{0}^{2}/I,
\label{startgl}
\end{equation}
is proportional to the square of the decay rate $\Gamma_{0}$ of the lasing
cavity mode and inversely proportional to the output power $I$ (in
units of photons/s). This is a lower bound for the linewidth when 
$\Gamma_{0}$ is much less than the linewidth of the atomic transition and when the lower level of the transition is
unoccupied (complete population inversion). While Schawlow and Townes had $K=1$, appropriate for a nearly closed cavity, it was later realized \cite{Pet79,Sie89} that an open cavity has an enhancement factor $K\geq 1$ called the ``Petermann factor''.

The RMT of the Petermann factor was developed in Refs.\ \cite{Pat00,Fra00}. The factor $K$ is related to the nonunitary matrix $U$ of right eigenvectors of ${\cal H}$, by
\begin{equation}
K=(U^{\dagger}U)_{00}(U^{-1}U^{-1\dagger})_{00},\label{Kdef}
\end{equation}
where the index $0$ labels the lasing mode. (In the presence of time reversal symmetry, one may choose $U^{-1}=U^{T}$, hence $K=[(UU^{\dagger})_{00}]^{2}$.) If the cavity is weakly coupled to the outside, then the matrix $U$ is unitary and $K=1$, but more generally $K\geq 1$. The probability distribution $P(K|\Gamma_{0})$ of the Petermann factor for a given value of the decay rate $\Gamma_{0}$ is very broad and asymmetric, with an algebraically decaying tail towards large $K$. For example, in the case $N=1$ of a single-mode opening of the cavity, $P(K|\Gamma_{0})\propto (K-1)^{-2-3\beta/2}$.

\acknowledgments

I have received helpful feedback on this article from Y. Alhassid, B. B\'{e}ri, H. Schomerus, J. Schliemann, A. D. Stone, and M. Titov. I am also indebted to the authors of the publications from which I reproduced figures, for their permission. My research is funded by the Dutch Science Foundation NWO/FOM and by the European Research Council.

%\newpage

\end{document}